\documentclass[showpacs,amsmath,twocolumn,showkeys,pra]{revtex4-1}
\usepackage{dcolumn}    
\usepackage{bm}         
\usepackage{ifpdf}
\usepackage{amssymb,lineno,amsfonts}
\usepackage{graphicx}   
\usepackage{bbm}
\usepackage{mathrsfs}
\usepackage{upgreek}
\usepackage{mathtools}
\usepackage{epstopdf}
\usepackage{setspace}
\usepackage{hyperref}
\usepackage[matrix,frame,arrow]{xypic}
\usepackage{rotating}
\usepackage{float}
\usepackage{natbib}
\usepackage{color} 
\definecolor{med-blue}{RGB}{25,25,112} 
\hypersetup{colorlinks, linkcolor={red},citecolor={blue}, urlcolor={blue}}

\newcommand{\ket}[1]{\vert{#1}\rangle}

\newcommand{\outpr}[2]{\vert{#1}\rangle\langle{#2}\vert}

\begin{document}
\title{Single-Shot Quantum Process Tomography}
\author{Abhishek Shukla and T. S. Mahesh}
\email{mahesh.ts@iiserpune.ac.in}
\affiliation{Department of Physics and NMR Research Center,\\
Indian Institute of Science Education and Research, Pune 411008, India}

\begin{abstract}
{
The standard procedure for quantum process tomography (QPT) involves
applying the quantum process on a system initialized in each of
a complete set of orthonormal states.  The corresponding outputs are then
characterized by quantum state tomography (QST), which
itself requires the measurement of non-commuting observables realized by independent experiments on 
identically prepared system states.  
Thus QPT procedure demands a number of independent measurements, and moreover, 
this number increases rapidly with the size of the system.
However, the total number of independent measurements can be greatly reduced 
with the availability of ancilla qubits.
Ancilla assisted process tomography (AAPT) has earlier been shown to require
a single QST of system-ancilla space.
Ancilla assisted quantum state tomography (AAQST) has also been shown to
perform QST in a single measurement.
Here we combine AAPT with AAQST to realize a `single-shot QPT' (SSPT), a procedure to characterize a general quantum
process in a single collective measurement of a set of commuting observables.
We demonstrate experimental SSPT by characterizing several single-qubit processes using
a three-qubit NMR quantum register. Furthermore, using the SSPT procedure
we experimentally characterize the twirling process and compare the results
with theory.}
\end{abstract}

\maketitle

\section{Introduction}
An open quantum system may undergo an evolution
due to intentional control fields as well as due to
unintentional interactions with stray fields caused by environmental
fluctuations.  Even the carefully designed control fields may
be imperfect to the extent that one might need to characterize
the overall process acting on the quantum system.  Such a
characterization, achieved by a procedure called quantum
process tomography (QPT), is crucial in the physical realization of a
fault-tolerant quantum processor  \cite{chuang97,zollar97}. 
QPT is realized by considering the quantum process as a map from
a complete set of initial states to final states, and experimentally
characterizing each of the final states using quantum state tomography (QST) \cite{ChuangPRL98}.
Since the spectral decomposition of a density matrix may involve
noncommuting observables,
Heisenberg's uncertainty principle demands multiple experiments to 
characterize the quantum state.
Thus QST by itself involves the measurement of a series of observables 
after identical preparations of the system in the quantum state. 
Hence, QPT in general requires a number of independent 
experiments, each involving initialization of the quantum system, 
applying the process to be characterized, and finally QST.
Furthermore, the total number of independent measurements required for QPT increases 
exponentially with the size of the system undergoing the process.

The physical realization of QPT has been demonstrated on various 
experimental setups such as NMR \cite{ChuangPRA2001,QPTofQFT}, linear optics \cite{EAPT1Exp,altepeter,obrian,bellstatefilter},
ion traps \cite{qptITprl2006,QPT_IonTrapNature2010}, superconducting qubits
\cite{QPT_SQUID,SQUID2009,MartiniSQUID2010,QPT2spSQUID,QPT_SQUIDChow2011,SQUID_Dewes2012}, and
NV center qubit \cite{suterprotectedgate}.  
Several developments in the methodology of QPT have also been
reported \cite{simplifiedQPT1,simplifiedQPT2}.  In particular, it has been shown that ancilla assisted 
process tomography (AAPT) can characterize a process with a single QST
\cite{mazzei2003pauli,EAPT1Exp,PRLAriano2003,altepeter}.
However, it still requires multiple
measurements each taken over a set of commuting observables. 
On the other hand, if sufficient ancilla qubits are available, QST can be
carried out with a single joint measurement over the entire system-ancilla space.
This procedure, known as ancilla assisted quantum state tomography
(AAQST), has been studied both theoretically and experimentally
\cite{Allahverdyan,Suteraaqst,peng,abhishek}.
Here we combine AAPT with AAQST and realize a `single-shot quantum process tomography' 
(SSPT), which can characterize a general process in a single collective measurement 
of the system-ancilla state.

In the next section, after briefly revising QPT and AAPT, we describe SSPT procedure.
In section III, we illustrate SSPT using a three-qubit nuclear magnetic resonance (NMR) 
quantum register.  We characterize certain unitary processes corresponding to 
standard quantum gates.  We also characterize a nonunitary process, namely twirling
operation. Finally we conclude in section IV.

 \begin{center}
\begin{figure}
\hspace*{-0.2cm}
\includegraphics[trim=0cm 0cm 0cm 0cm, clip=true,width=9cm]{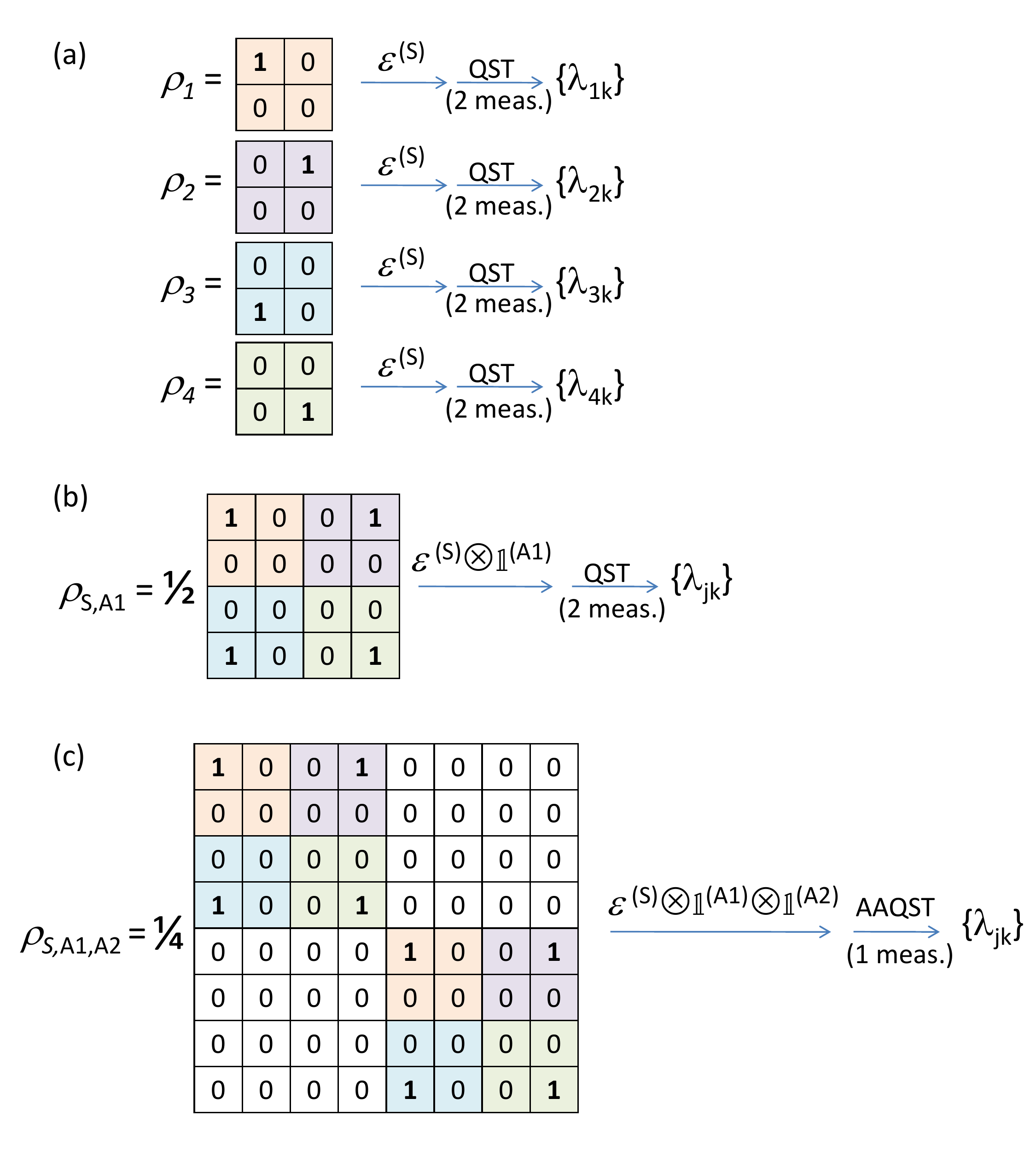} 
\caption{(Color online) Illustrating (a) single-qubit QPT 
requiring a total of 8 NMR measurements, (b) AAPT requiring 2 NMR measurements, and (c) SSPT
requiring a single NMR measurement. 
}
\label{sum} 
\end{figure}
\end{center}

\section{Theory}
\subsection{Quantum Process Tomography (QPT)}
A process $\varepsilon$ maps a quantum state $\rho$ 
to another state $\varepsilon(\rho)$.  Here we consider an $n$-qubit system with
$N^2( = 2^{2n})$-dimensional Liouville space $S$.
In order to characterize $\varepsilon$,
we let the process act on each linearly independent element of a complete basis set $\{\rho_1, \rho_2, \cdots, \rho_{N^2}\}$. 
Expressing each output state in the complete basis we obtain 
\begin{eqnarray}
\varepsilon(\rho_j) = \sum_k \lambda_{jk}\rho_k,
\label{lb}
\end{eqnarray}
where the complex coefficients $\lambda_{jk}$ can be extracted after QST.

The outcome of a trace-preserving quantum process $\varepsilon$ also has an operator-sum representation 
\begin{eqnarray}
\varepsilon(\rho) = \sum_i E_i \rho E_i^\dagger,
\end{eqnarray}
where the \textit{Kraus operators} $E_i$ satisfy the completeness relation $\sum_i E_i^\dagger E_i = I$.
To assist experimental characterization of the process, we utilize 
a fixed set of basis operators $\{\tilde{E}_m\}$, and express 
$E_i = \sum_m e_{im} \tilde{E}_m$.
The process is now described by
\begin{eqnarray}
\varepsilon(\rho) = \sum_{mn}\tilde{E}_m \rho \tilde{E}_n^\dagger \chi_{mn},
\label{tildeE}
\end{eqnarray}
where $\chi_{mn} = \sum_i e_{im} e_{in}^*$ form a complex matrix which completely characterizes the process $\varepsilon$.
Since the set $\{\rho_k\}$ forms a complete basis, it is also possible to express
\begin{eqnarray}
\tilde{E}_m\rho_j\tilde{E}_n^\dagger = \sum_k \beta_{jk}^{mn}\rho_k,
\label{beta}
\end{eqnarray}
where $\beta_{jk}^{mn}$ can be calculated theoretically. Eqns. \ref{lb}, \ref{tildeE} and \ref{beta}
lead to
\begin{eqnarray}
\varepsilon(\rho_j) =  \sum_k \lambda_{jk}\rho_k  = \sum_k \sum_{mn} \beta_{jk}^{mn} \chi_{mn} \rho_k.
\end{eqnarray}
Exploiting the linear independence of $\{\rho_k\}$, one obtains the matrix equation
\begin{eqnarray}
\beta \chi = \lambda,
\label{chi}
\end{eqnarray}
from which $\chi$-matrix can be extracted by standard methods in linear algebra.

For example, in the case of a single qubit, one can choose the linearly independent basis
$\{\outpr{0}{0},\outpr{0}{1},\outpr{1}{0},\outpr{1}{1}\}$ (see Fig. \ref{sum}a). 
While the middle-two elements are non-Hermitian, they
can be realized as a linear combination of Hermitian density operators \cite{chuangbook}.
A fixed set of operators $\{I,X,-iY,Z\}$ can be used to express the
$\chi$ matrix.  Thus the standard single-qubit QPT procedure requires four QST experiments.

QPT on an $N$-dimensional system requires $N^2$-QST experiments, and
a single QST involves several quantum measurements each taken jointly over a set of
commuting observables.  The exact number of measurements may depend on the
properties of available detectors. 
In the case of an $n$-qubit NMR system with a well resolved spectrum \cite{cavanagh},
QST requires $\simeq \left\lceil \frac{N}{n} \right\rfloor$
measurements, where $\lceil \rfloor$ rounds the argument to next integer \cite{abhishek}.
Therefore an $n$-qubit QPT needs a total of $M_\mathrm{QPT} \simeq N^2 \left\lceil \frac{N}{n} \right\rfloor$ measurements. 
Estimates of $M$ for a small number of qubits
shown in the first column of Table 1 illustrate the exponential increase of $M_\mathrm{QPT}$ with $n$.

\begin{table}
$
\begin{array}{|c|c|cc|cc|}
           \hline
           n & M_\mathrm{QPT} & M_\mathrm{AAPT} & (n_{A1})& M_\mathrm{SSPT} &(n_{A1}, n_{A2})\\
           \hline
           \hline
           1      &     8     &      2 & (1)     &     1 &(1,1) \\
           \hline
           2    &      32      &     4 &(2)     &     1 &(2,2) \\
           \hline
           3    &     192      &    11 &(3)     &     1 &(3,3) \\
           \hline
           4    &    1024      &    32 &(4)     &     1 &(4,5) \\
           \hline
           5   &     7168    &     103 &(5)     &     1 &(5,6) \\
           \hline
\end{array}
$
\caption{Comparison of number of independent measurements and 
number of ancilla qubits (in parenthesis) required for 
$n$-qubit QPT, AAPT, and SSPT.}
\end{table}

\subsection{Ancilla-Assisted Process Tomography (AAPT)}
If sufficient number of ancillary qubits are available, 
ancilla assisted process tomography (AAPT) can be carried out
by simultaneously encoding all the basis elements onto a higher 
dimensional system-ancilla Liouville space $S \otimes A1$ \cite{mazzei2003pauli,EAPT1Exp,PRLAriano2003,altepeter}.
AAPT requires a single final QST, thus greatly reducing the number of independent measurements.
For example, a single-qubit process tomography can be carried out with the
help of an ancillary qubit by preparing the $n-$qubit cat state 
$\ket{\psi_n} = (\ket{00 \cdots 0}+\ket{11 \cdots 1})/\sqrt{2}$,
applying the process on the system-qubit, and finally carrying out QST of the
two-qubit state (see Fig. \ref{sum}b). Although  
only two independent measurements are needed for a two-qubit QST, this number grows exponentially with the
total number of qubits.

All the $N^2$ basis elements of the $n$-qubit system
can be encoded simultaneously in independent subspaces of a single $N^2 \times N^2$ Liouville operator 
belonging to $2n$-qubit space $S \otimes A1$.  Thus exactly $n$-ancilla qubits are needed to carry out AAPT
on an $n$-qubit system.  
Therefore AAPT requires $M_{\mathrm{AAPT}}\simeq \left\lceil \frac{N^2}{2n} \right\rfloor$ 
independent measurements.  The minimum number of experiments for a few system-qubits 
are shown in the second column of Table I. While AAPT requires significantly
lesser number of measurements compared to QPT, it still scales exponentially with
the number of system-qubits.

\subsection{Single-Shot Process Tomography (SSPT)}
It had been shown earlier that, if sufficient number 
of ancillary qubits are available, QST of a general density matrix of
arbitrary dimension can be performed with a single joint measurement 
of a set of commuting observables \cite{Allahverdyan,Suteraaqst,peng,abhishek}.  
This method, known as
ancilla assisted quantum state tomography (AAQST) is based on the redistribution
of all elements of the system density matrix on to a joint density matrix
in the combined system-ancilla Liouville space.
Initially ancilla register for AAQST is prepared in a maximally mixed state thus erasing all information in it
and redistribution of matrix elements is achieved by an optimized joint unitary operator \cite{abhishek}.

By combining AAPT with AAQST, process tomography can be achieved with
a single joint measurement of all the qubits (see Fig. \ref{sum}c  and 3rd
column of Table I).
If AAQST is carried out with an ancilla space ($A2$) of $n_{A2}$-qubits, the combined space
$S \otimes A1 \otimes A2$ corresponds to $\tilde{n} = 2n+n_{A2}$ qubits.
A single joint measurement suffices if the total number of observables 
is equal to or exceeds the number of real unknowns (i.e., $N^4-1$) in the $2n$-qubit density matrix, i.e.,
if $\tilde{n}{\tilde{N}} \ge  (N^4-1)$,
where $\tilde{N} = 2^{\tilde{n}}$ \cite{abhishek}.
The number of ancillary qubits $n_{A1}$ and $n_{A2}$ required for SSPT are shown in the third column of
Table I.

\begin{figure}
 \begin{center}
\hspace*{-0.5cm}
\includegraphics[trim=2.5cm 5cm 3.5cm 3cm, clip=true,width=8.5cm]{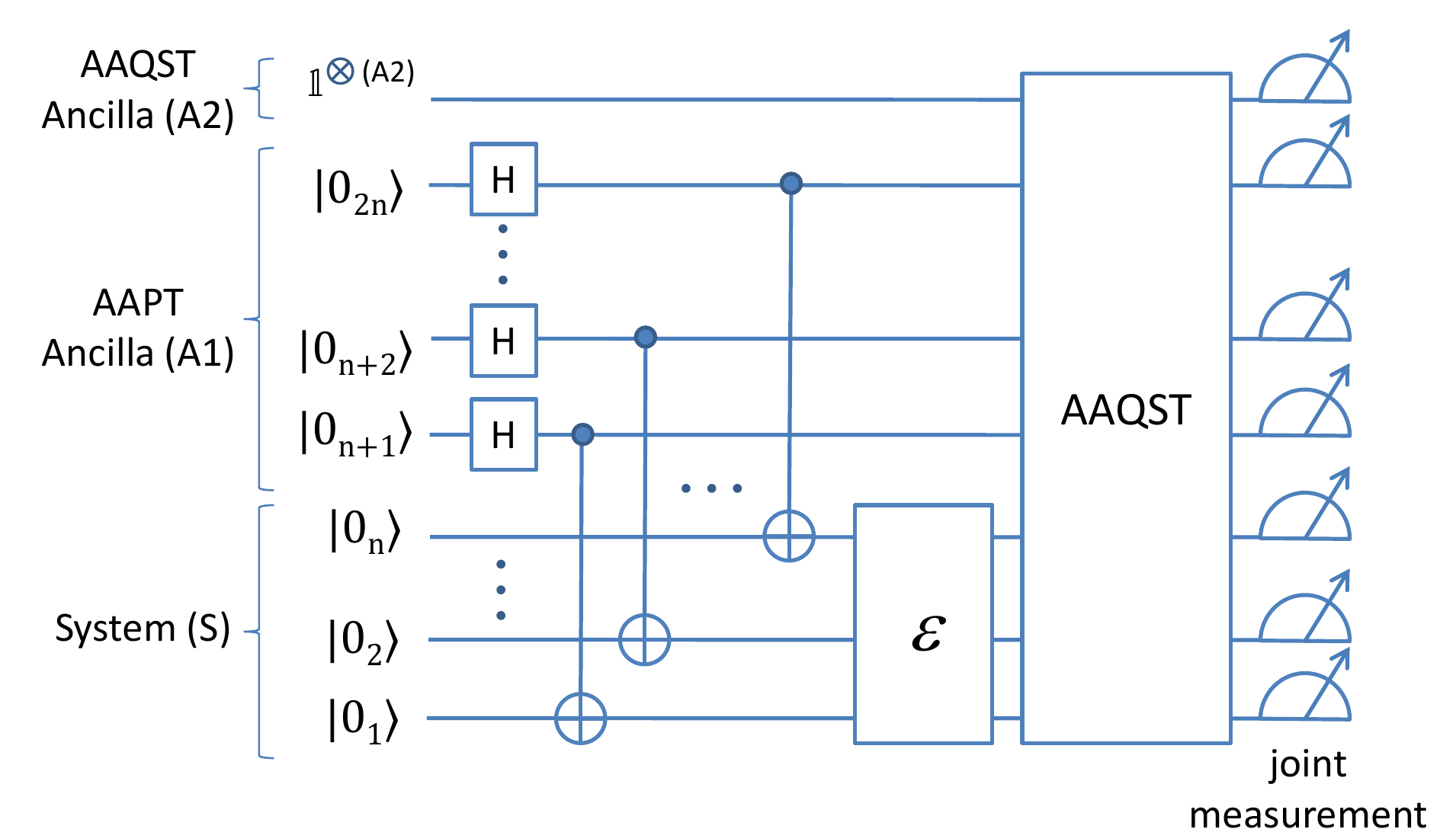} 
\caption{(Color online) 
Quantum circuit for SSPT.
}
\label{qckt} 
\end{center}
\end{figure}

The complete circuit for SSPT is shown in Fig. \ref{qckt}. It involves two ancilla
registers, one for AAPT and the other for AAQST.  Initially AAQST register is
prepared in a maximally mixed state and the other two registers are set to $\ket{0^{\otimes{n}}}$ states.  
Hadamard gates on the AAPT ancilla followed by C-NOT gates (on system qubits
controlled by ancilla) prepare state $\ket{\psi_n}$, which simultaneously 
encodes all the basis elements required for QPT. A single application of the process $\varepsilon$, on the system
qubits, acts simultaneously and independently on all the basis elements $\{\rho_j\}$. 
The final AAQST operation allows estimation of all the elements of the $2n$-qubit
density matrix $\sum_j \varepsilon(\rho_j)\otimes A1^{(j)}$, 
where $A1^{(j)}$ identifies the $j$th subspace.
The output of each subspace $\varepsilon(\rho_j)$ can now be extracted, and the coefficients $\lambda_{jk} = Tr[\varepsilon(\rho_j) \rho_k^\dagger]$ can be calculated.

\begin{figure}[b]
 \begin{center}
\hspace*{-0.5cm}
\includegraphics[trim=0cm 2cm 0cm 2cm, clip=true,width=9.6cm]{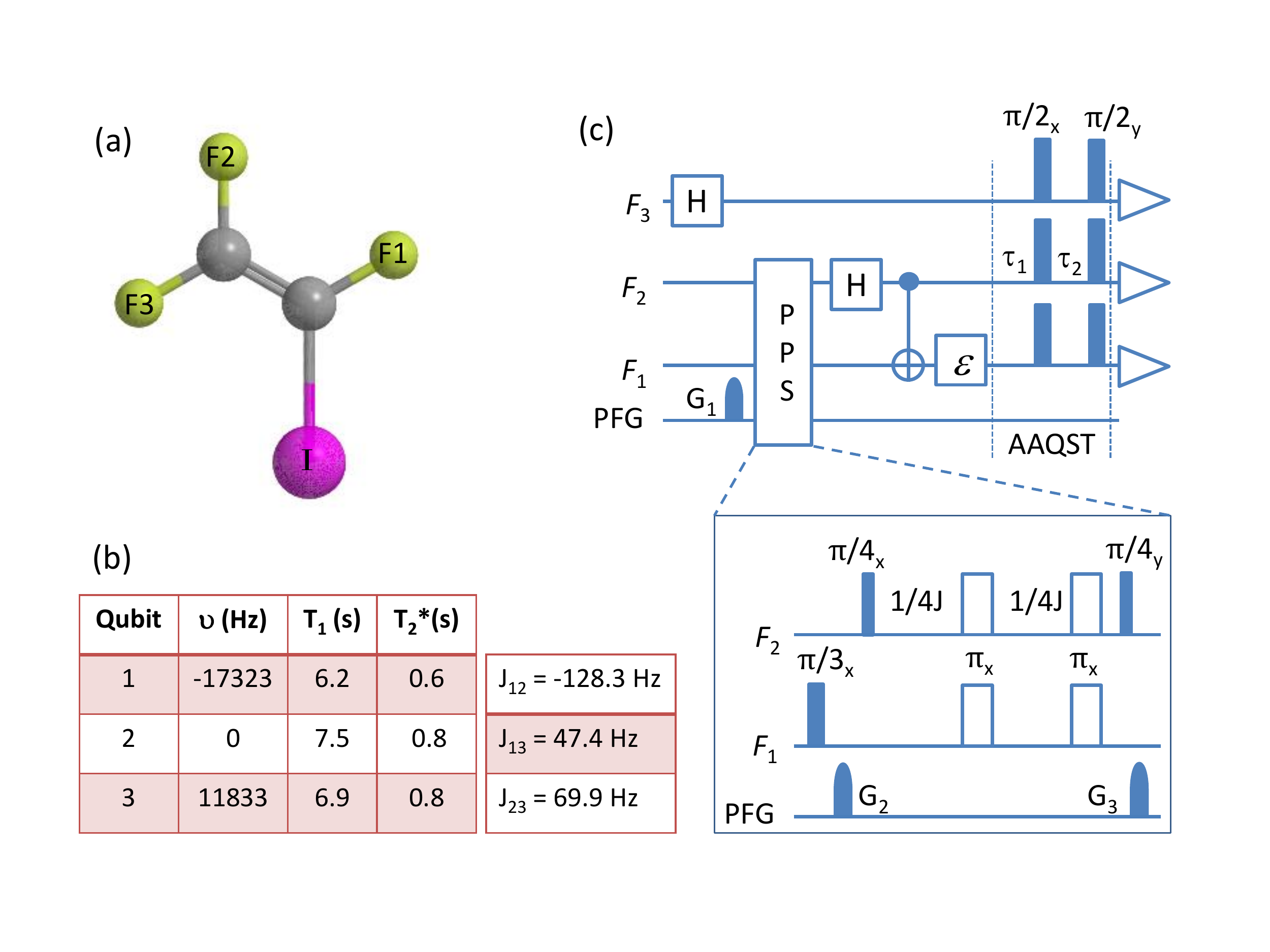} 
\caption{(Color online) Molecular structure of iodotrifluoroethylene (a), and
the table of Hamiltonian and relaxation parameters (b), NMR pulse-sequence 
to demonstrate SSPT (c). Pulse sequence for preparing $\ket{00}$ pseudopure state
is shown in the inset of (c). \label{fffmol} }
\end{center}
\end{figure}

\section{Experiments}
We used iodotrifluoroethylene (C$_2$F$_3$I) dissolved in acetone-D$_6$
as a 3-qubit system.  The molecular structure and labelling scheme are shown in Fig. \ref{fffmol} (a).  
All the experiments described below are carried out
on a Bruker 500 MHz NMR spectrometer at an ambient temperature of 300 K using high-resolution
NMR techniques.  
The NMR Hamiltonian in this case can be expressed as
\begin{eqnarray}
{\cal H} = -\pi\sum_{i=1}^{3}\nu_i \sigma_z^i  + 
\pi\sum_{i=1,j > i}^{3,3}  J_{ij} \sigma_z^i \sigma_z^j /2
\label{ham}
\end{eqnarray}
where $\sigma_{z}^i$ and $\sigma_{z}^j$ are Pauli $z$-operators
of $i$th and $j$th qubits \cite{cavanagh}. The chemical shifts $\nu_i$,
coupling constants $J_{ij}$, and relaxation parameters (T$_1$ and T$_2^*$)
are shown in Fig.  \ref{fffmol} (b).
All the pulses are realized using gradient ascent pulse engineering (GRAPE)
technique \cite{khaneja2005optimal} 
and had average fidelities above 0.99 over 20\% inhomogeneous
RF fields.

We utilize spins F$_1$, F$_2$, and F$_3$ respectively as
the system qubit ($S$), AAPT ancilla ($A_1$), and AAQST ancilla ($A_2$). 
The NMR pulse-sequence for SSPT experiments are shown in Fig. \ref{fffmol}(c).
It begins with preparing A2 qubit in the maximally mixed state by
bringing its magnetization into transverse direction using a Hadamard gate,
and subsequently dephasing it using a PFG. 
The remaining qubits are initialized into a pseudopure  $\ket{00}$ state
by applying the standard pulse-sequence shown in the inset of Fig. \ref{fffmol}c \cite{cory2000nmr}.
The Bell state $\ket{\psi_2}$ prepared using a
Hadamard-CNOT combination had a fidelity of over 0.99. 
After preparing this state, we applied the process $\varepsilon$ on 
the system qubit. The final AAQST consists of $(\pi/2)_x$ and $(\pi/2)_y$
pulses on all the qubits separated by delays $\tau_1 = 6.7783$ ms and $\tau_2 = 8.0182$ ms
\cite{abhishek}. A single joint measurement of all the qubits now leads to a complex
signal of 12 transitions, from which all the 15 real unknowns of the
2-qubit density matrix $\rho_{S,A1} = \sum_j \varepsilon(\rho_j)\otimes A1^{(j)}$ of $F_1$ and $F_2$
can be estimated \cite{abhishek}.  In our choice of fixed set of operators and basis elements
\begin{eqnarray}
\rho_{S,A1} = \left[
\begin{array}{cc|cc}
\lambda_{11} & \lambda_{12} & \lambda_{21} & \lambda_{22} \\
\lambda_{13} & \lambda_{14} & \lambda_{23} & \lambda_{24} \\
\hline
\lambda_{31} & \lambda_{32} & \lambda_{41} & \lambda_{42} \\
\lambda_{33} & \lambda_{34} & \lambda_{43} & \lambda_{44}
\end{array}
\right] 
\label{rhosa1}
\end{eqnarray}
\cite{sup}.
The $\chi$ matrix characterizing the complete process can now be obtained
by solving the eqn. \ref{chi}.

\begin{figure}
\includegraphics[trim=9.5cm 0cm 0cm 1cm, clip=true,width=8.8cm]{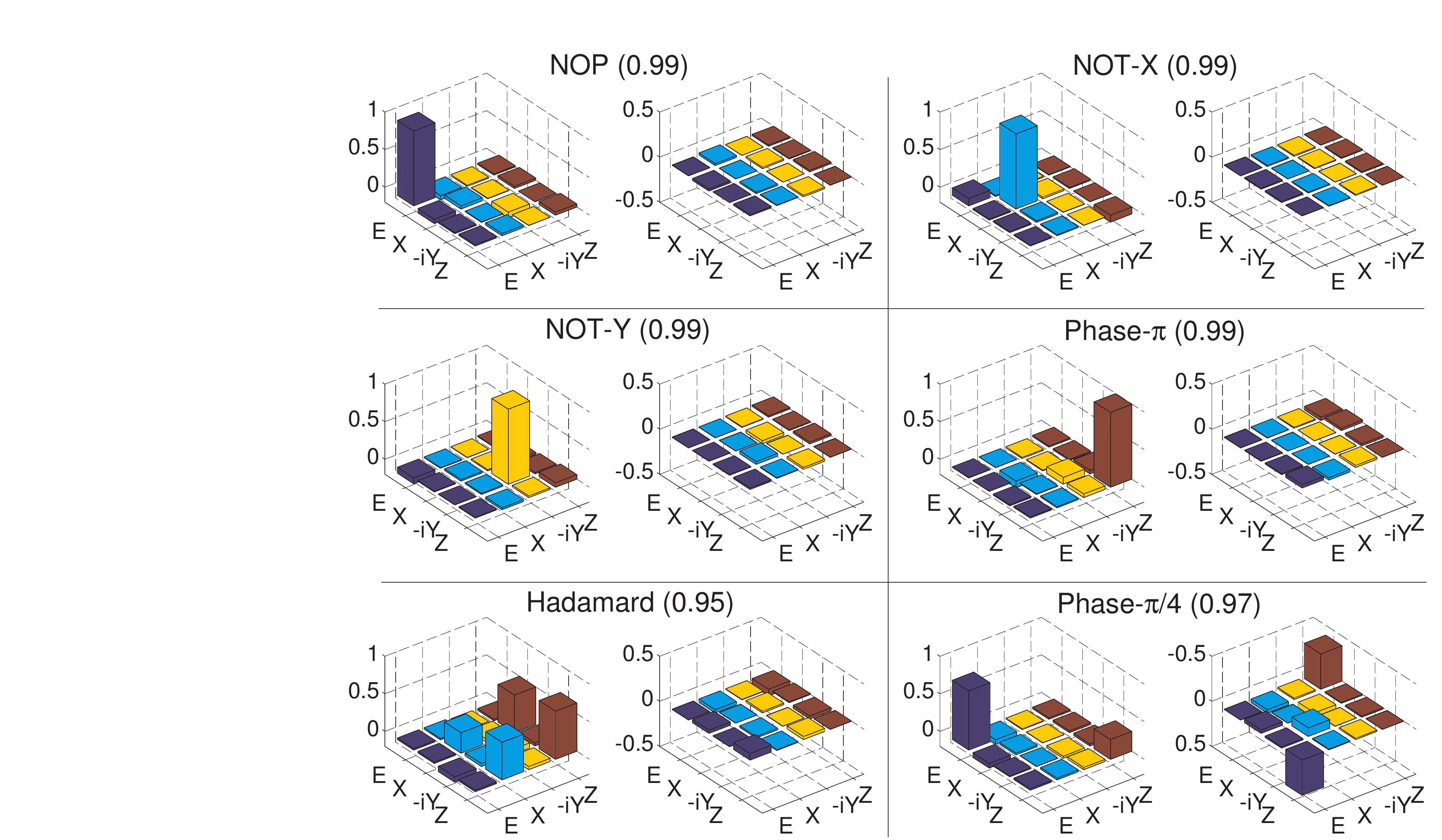} 
\caption{(Color onlinee) The barplots showing experimental $\chi$-matrices 
for various quantum processes obtained using SSPT.  In each case, the left and right barplots
correspond to the real and imaginary parts respectively, and the fidelities are indicated
in parenthesis.
\label{ssptres}
} 
\end{figure}

\subsection{SSPT of quantum gates}
We now describe experimental characterization of several single-qubit unitary processes.
The quantum gates to be characterized are introduced as process $\varepsilon$ on
F$_1$ qubit in Fig. \ref{fffmol}c.
The experimental $\chi$-matrices for 
NOP (identity process), NOT-X ($e^{-i \pi X/2}$), NOT-Y ($e^{-i \pi Y/2}$), 
Hadamard, Phase$-\pi$ ($e^{i \pi Z}$), and Phase$-\pi/4$ ($e^{i \pi Z/8}$) are
shown in Fig. \ref{ssptres}. 
Starting from thermal equilibrium, each SSPT experiment
characterizing an entire one-qubit process took less than four seconds.
A measure of overlap of the experimental process $\chi_\mathrm{exp}$
with the theoretically expected process $\chi_\mathrm{th}$ is given
by the gate fidelity \cite{suterprotectedgate}
\begin{eqnarray}
F(\chi_\mathrm{exp},\chi_\mathrm{th}) = \frac{\vert Tr[\chi_\mathrm{exp} \chi_\mathrm{th}^\dagger] \vert}
{\sqrt{Tr[\chi_\mathrm{exp}^\dagger \chi_\mathrm{exp}] ~ Tr[\chi_\mathrm{th}^\dagger \chi_\mathrm{th}]}}.
\label{gatefid}
\end{eqnarray}
The gate fidelities for all the six processes are indicated
in Fig. \ref{ssptres}.
Except in the cases of Hadamard and
Phase-$\pi/4$, the gate fidelities were about 0.99.  
The lower fidelities in Hadamard (0.95) and Phase-$\pi/4$ (0.97) are
mainly due to imperfections in the RF pulses implementing these processes.

\begin{figure}[b]
\begin{center}
\includegraphics[trim=0cm 1cm 0cm 0cm, clip=true,width=7cm]{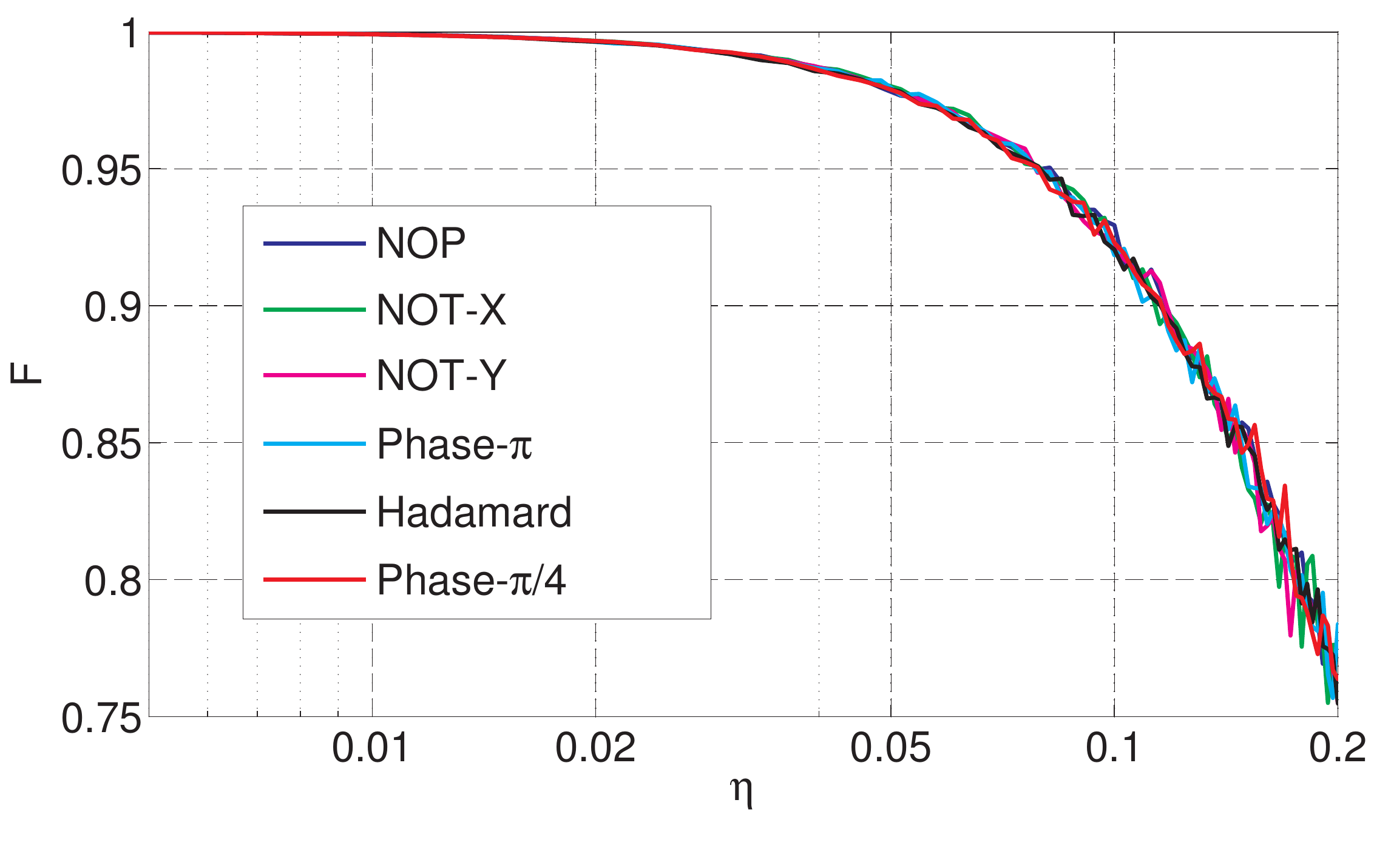} 
\caption{(Color online) 
Simulated fidelity of various processes as a function of noise $\eta$.
\label{ssptrob} 
}
\end{center}
\end{figure}

In order to study the robustness of SSPT procedure we first considered an ideal
process, simulated the corresponding spectral intensities, and reconstructed
the final density matrix $\rho_{S,A1}$. 
Using eqn. \ref{rhosa1} we obtained $\lambda_{jk}$ values and calculated
the  matrix $\chi_0$ simulating the noise-free SSPT procedure.
We then introduced noise by adding random numbers in the range $[-\eta,\eta]$ to the spectral intensities 
and used the resulting data calculating $\chi_\eta$ simulating the noisy SSPT procedure.
The variations of average gate fidelities $F(\chi_0,\chi_\eta)$ for various processes versus noise amplitude
$\eta$ are shown in Fig. \ref{ssptrob}.  Interestingly, the noise has similar effects on
fidelities of all the simulated processes.  We also observe that fidelities remained
above 0.9 for $\eta < 0.1$, indicating that SSPT is fairly robust against the noise
in this range.

\subsection{SSPT of twirling process}
Twirling is essentially a nonunitary process usually realized by an ensemble average
of a set of unitary operations. It was first introduced
by Bennett et al \cite{benett1} for extracting singlet states from 
a mixture of Bell states.  Twirling has been studied in detail 
\cite{benett2,emerson1,emerson2,emerson3,corytwirl}
and various modified twirling protocols have also been suggested \cite{dankert2009,laflammeprl2012}.

In NMR, twirling can be achieved with the help of a pulsed field gradient (PFG), which 
produces a continuous space-dependent spin-rotation, such that the ensemble average
effectively emulates a nonunitary process \cite{anwar}. A $\hat{z}$ PFG produces a 
space-dependent unitary $U_\phi = \exp\left(-i \frac{\phi}{2}\sum_{j=1}^n \sigma_{jz}\right)$ 
with $\phi \in [-\Phi,\Phi]$, where $j$ is the summation index over all the qubits, and 
the maximum phase $\Phi$ depends on the strength and duration of PFG
and the sample volume.  
When the $\hat{z}$ PFG acts on an initial $n$-qubit density matrix $\rho_{\mathrm{in}} = \sum_{lm} \rho_{lm}\outpr{l}{m}$,
the resultant output density matrix is,
\begin{eqnarray}
\rho_{\mathrm{out}} 
&=& \frac{1}{2\Phi}\int_{-\Phi}^{\Phi} d\phi 
~U_\phi \rho_{\mathrm{in}} U_\phi^\dagger
\nonumber \\
&=&  \sum_{lm} \rho_{lm}\outpr{l}{m} ~ \mathrm{sinc}(q_{lm}\Phi).
\label{twirlrho}
\end{eqnarray}
Here $\mathrm{sinc(x)} = \frac{\sin x}{x}$ and $q_{lm} = \frac{1}{2} \sum_j \left[ (-1)^{m_j} - (-1)^{n_j} \right]$ is
the \textit{quantum number} of the element $\outpr{l_1 l_2 \cdots l_n}{m_1 m_2 \cdots m_n}$, i.e., the
difference in the spin-quantum numbers of the corresponding basis states.
While the diagonal elements $\outpr{l}{l}$ 
and other zero-quantum elements are unaffected by twirling, 
the off-diagonal elements with $q_{lm} \neq 0$ undergo decaying oscillations with increasing $\Phi$ values.

SSPT of twirling process is carried out using the procedure described in
Fig. \ref{fffmol}c after introducing $\delta$-PFG-$\delta$ in place of the process $\varepsilon$,
where $\delta$ is a short delay for switching the gradient.
Applying PFG selectively on the system qubit is not simple, and is also unnecessary.
Since the F$_3$ qubit (AAQST ancilla) is already in a maximally mixed state, twirling has no effect on it.
For the Bell state $\ket{\psi_2}$, applying a strong twirling on either or both spins (F$_1$, F$_2$) has the same
effect, i.e., a strong measurement  reducing the joint-state to a 
maximally mixed state. However, since $\ket{\psi_2}$ corresponds to a two-quantum 
coherence (i.e., $q_{00,11}=2)$, its dephasing is double that of a single-quantum coherence.
Assuming the initial state $\rho_{\mathrm{in}} = \outpr{\psi_2}{\psi_2}$, and
using expressions \ref{lb} and \ref{twirlrho}, we find that the non-zero elements of $\lambda$ are
\begin{eqnarray}
\lambda_{11} = \lambda_{44} = 1, ~ \mathrm{and,} ~ \lambda_{22} = \lambda_{33} = \mathrm{sinc}(2\Phi).
\end{eqnarray}
Solving expression \ref{chi}, we obtain a real $\chi$ matrix with only nonzero elements 
\begin{eqnarray}
\chi_\mathrm{EE} = \frac{1+\mathrm{sinc}(2\Phi)}{2} ~ \mathrm{and} ~  
\chi_\mathrm{ZZ} =  \frac{1-\mathrm{sinc}(2\Phi)}{2}.
\label{chith}
\end{eqnarray}

\begin{figure}[t]
\begin{center}
\includegraphics[trim=3.5cm 0cm 0cm 1.8cm, clip=true,width=8.9cm]{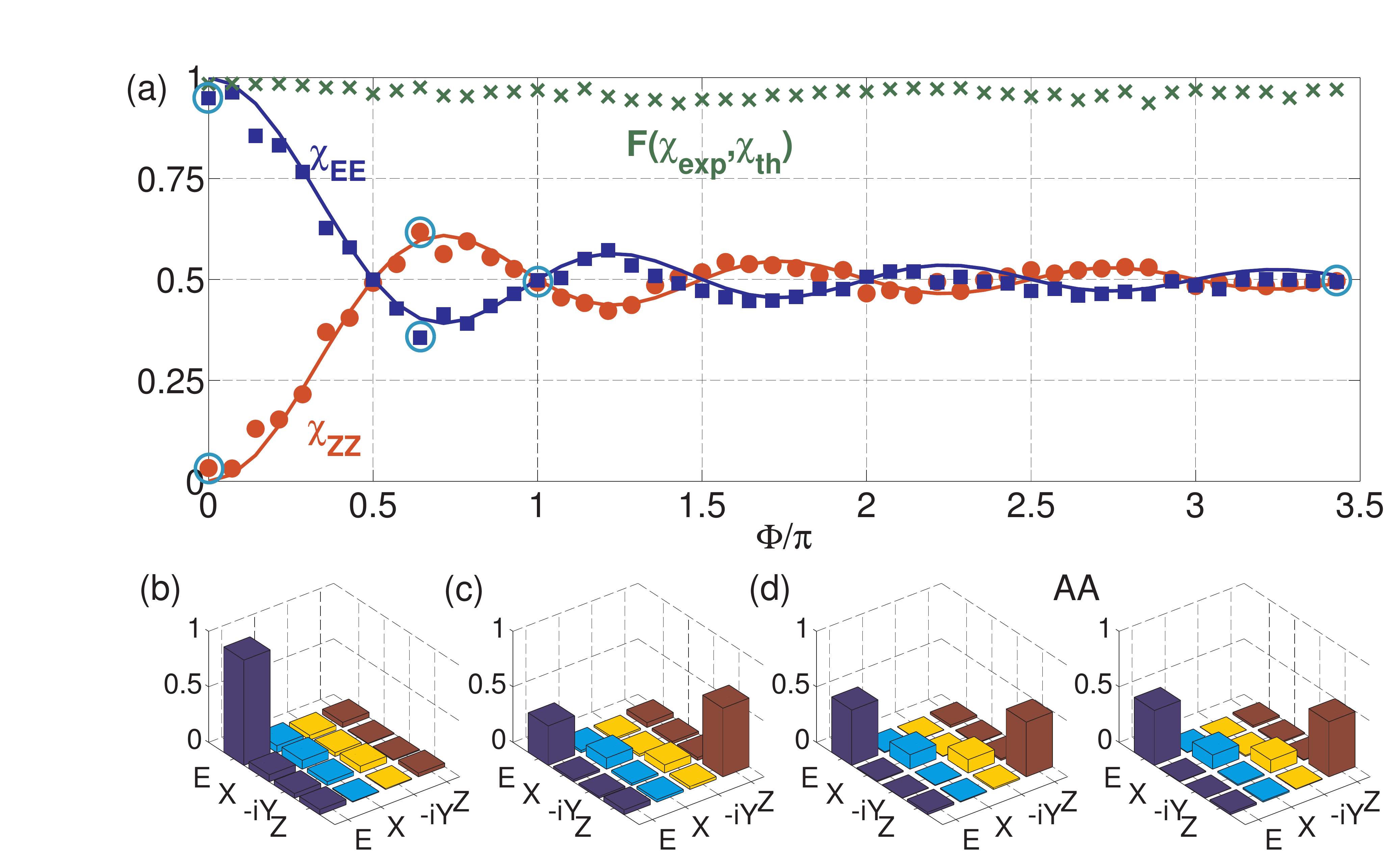} 
\caption{(Color online) 
(a) The experimental values of $\vert \chi_{\mathrm{EE}} \vert$ 
($\vert \chi_{\mathrm{ZZ}} \vert$) are shown
by filled squares (filled circles).  
The solid line ($\chi_{\mathrm{EE}}$) and the dashed line
($\chi_{\mathrm{ZZ}}$) illustrate theory.
The barplots correspond to experimental $\vert \chi \vert$ matrices at 
(b) $\Phi = 0$, (c) $\Phi = 0.64\pi$, (d) $\Phi = \pi$, and (e) $\Phi = 3.43 \pi$.
\label{twirlres} 
}
\end{center}
\end{figure}

In our experiments, the duration of PFG and $\delta$ are set to 300 $\upmu$s 
and 52.05 $\upmu$s respectively, 
such that the chemical shift evolutions and J-evolutions are negligible.  The strength of twirling
is slowly varied by increasing the PFG strength from 0 to 2.4 G/cm in steps of 0.05 G/cm.
The results of the experiments are shown in Fig. \ref{twirlres}.
The filled squares (circles) in Fig. \ref{twirlres}(a) correspond to experimentally
obtained values for $\vert \chi_{\mathrm{EE}}\vert$ ($\vert \chi_{\mathrm{ZZ}}\vert$).
Small imaginary parts observed in experimental $\chi$ matrices are due to minor experimental imperfections.
The smooth (dashed) lines indicate corresponding theoretical values obtained from eqns. \ref{chith}.
The crosses indicate the gate fidelities $F(\chi_\mathrm{exp},\chi_\mathrm{th})$ 
calculated using eqn \ref{gatefid}.
The barplots show experimental $\vert \chi \vert$ matrices for 
(b) $\Phi = 0$, (c) $\Phi = 0.64\pi$, (d) $\Phi = \pi$, and (e) $\Phi = 3.43 \pi$, and
$\chi_{\mathrm{EE}}$ and $\chi_{\mathrm{ZZ}}$ values in Fig \ref{twirlres}(a)  
corresponding to these $\Phi$ values are circled out.

At zero twirling, the process is essentially a NOP process as is clear
from the bar plot in Fig. \ref{twirlres}(b), wherein $\vert \chi_{\mathrm{EE}} \vert \approx 1$ 
and $\vert \chi_{\mathrm{ZZ}} \vert \approx 0$. 
When $\Phi = k \pi/2$ with an integer $k$, the ensemble initially prepared in state $\ket{\psi_2}$ undergoes
an overall phase distribution over $[-k\pi,k\pi]$, and at this stage $\chi_{\mathrm{EE}} =  \chi_{\mathrm{ZZ}} = 0.5$
(eg. Fig. \ref{twirlres} d).  Further increase in $\Phi$ leads to oscillations 
of $ \chi_{\mathrm{EE}} $ and $ \chi_{\mathrm{ZZ}} $ about 0.5, and for large 
$\Phi$ values, both of these elements damp towards 0.5 and all other elements
vanish (eg. Fig. \ref{twirlres} e).
The errors in experimental $\chi_\mathrm{EE}$ and $\chi_\mathrm{ZZ}$ values
were less than 8 \%.  The good agreement of the experimental values with theory
indicates the overall success of SSPT procedure.  The average of the gate fidelities 
was over 0.96. Small deviations of the experimental
values from theory are due to nonlinearities in PFG profile as well as
due to imperfections in RF pulses implementing the SSPT procedure.  

\section{Conclusions}
Information processing requires two important physical resources, namely, the size of the register
and the number of operations. Often there
exists an equivalence between these two resources which allows trading  one resource with another.
Likewise, in the present work we show that, if some extra qubits
are available, it is possible to carry out quantum process tomography of the
system qubits with a single joint measurement over a set of commuting observables. 
We have illustrated this method on a single system qubit and two ancillary
qubits using NMR quantum computing methods.  In particular, 
we extracted the $\chi$ matrices characterizing certain quantum gates and obtained their
gate fidelities with the help of a single measurement of a three qubit system in each case.
We studied the robustness of SSPT procedure using numerical simulations.
We also characterized twirling operation which is essentially a nonunitary process.
The overall procedure of SSPT is general and can be applied to other fields such as optical qubits,
trapped ions, or superconducting qubits.  

A potential application of single-shot
process tomography could be in high throughput characterization of dynamic processes.
The standard methods require repeated applications of the same process either to
collect independent outputs from all the basis states or to allow quantum state tomography.  
However, the present method requires only one application of the process for the
entire characterization.  

\section*{Acknowledgements}
The authors are grateful to Prof. Anil Kumar, 
Swathi Hegde, Hemant Katiyar, and Rama Koteswara Rao
for discussions. This work was partly supported by DST project SR/S2/LOP-0017/2009.

\bibliographystyle{apsrev4-1}
\bibliography{ref2}

\begin{thebibliography}{40}%
\makeatletter
\providecommand \@ifxundefined [1]{%
 \@ifx{#1\undefined}
}%
\providecommand \@ifnum [1]{%
 \ifnum #1\expandafter \@firstoftwo
 \else \expandafter \@secondoftwo
 \fi
}%
\providecommand \@ifx [1]{%
 \ifx #1\expandafter \@firstoftwo
 \else \expandafter \@secondoftwo
 \fi
}%
\providecommand \natexlab [1]{#1}%
\providecommand \enquote  [1]{``#1''}%
\providecommand \bibnamefont  [1]{#1}%
\providecommand \bibfnamefont [1]{#1}%
\providecommand \citenamefont [1]{#1}%
\providecommand \href@noop [0]{\@secondoftwo}%
\providecommand \href [0]{\begingroup \@sanitize@url \@href}%
\providecommand \@href[1]{\@@startlink{#1}\@@href}%
\providecommand \@@href[1]{\endgroup#1\@@endlink}%
\providecommand \@sanitize@url [0]{\catcode `\\12\catcode `\$12\catcode
  `\&12\catcode `\#12\catcode `\^12\catcode `\_12\catcode `\%12\relax}%
\providecommand \@@startlink[1]{}%
\providecommand \@@endlink[0]{}%
\providecommand \url  [0]{\begingroup\@sanitize@url \@url }%
\providecommand \@url [1]{\endgroup\@href {#1}{\urlprefix }}%
\providecommand \urlprefix  [0]{URL }%
\providecommand \Eprint [0]{\href }%
\providecommand \doibase [0]{http://dx.doi.org/}%
\providecommand \selectlanguage [0]{\@gobble}%
\providecommand \bibinfo  [0]{\@secondoftwo}%
\providecommand \bibfield  [0]{\@secondoftwo}%
\providecommand \translation [1]{[#1]}%
\providecommand \BibitemOpen [0]{}%
\providecommand \bibitemStop [0]{}%
\providecommand \bibitemNoStop [0]{.\EOS\space}%
\providecommand \EOS [0]{\spacefactor3000\relax}%
\providecommand \BibitemShut  [1]{\csname bibitem#1\endcsname}%
\let\auto@bib@innerbib\@empty
\bibitem [{\citenamefont {Chuang}\ and\ \citenamefont
  {Nielsen}(1997)}]{chuang97}%
  \BibitemOpen
  \bibfield  {author} {\bibinfo {author} {\bibfnamefont {I.~L.}\ \bibnamefont
  {Chuang}}\ and\ \bibinfo {author} {\bibfnamefont {M.~A.}\ \bibnamefont
  {Nielsen}},\ }\href@noop {} {\bibfield  {journal} {\bibinfo  {journal}
  {Journal of Modern Optics}\ }\textbf {\bibinfo {volume} {44}},\ \bibinfo
  {pages} {2455} (\bibinfo {year} {1997})}\BibitemShut {NoStop}%
\bibitem [{\citenamefont {Poyatos}\ \emph {et~al.}(1997)\citenamefont
  {Poyatos}, \citenamefont {Cirac},\ and\ \citenamefont {Zoller}}]{zollar97}%
  \BibitemOpen
  \bibfield  {author} {\bibinfo {author} {\bibfnamefont {J.~F.}\ \bibnamefont
  {Poyatos}}, \bibinfo {author} {\bibfnamefont {J.~I.}\ \bibnamefont {Cirac}},
  \ and\ \bibinfo {author} {\bibfnamefont {P.}~\bibnamefont {Zoller}},\ }\href
  {\doibase 10.1103/PhysRevLett.78.390} {\bibfield  {journal} {\bibinfo
  {journal} {Phys. Rev. Lett.}\ }\textbf {\bibinfo {volume} {78}},\ \bibinfo
  {pages} {390} (\bibinfo {year} {1997})}\BibitemShut {NoStop}%
\bibitem [{\citenamefont {I.~L.~Chuang}\ and\ \citenamefont
  {Leung}(1998)}]{ChuangPRL98}%
  \BibitemOpen
  \bibfield  {author} {\bibinfo {author} {\bibfnamefont {M.~G.~K.}\
  \bibnamefont {I.~L.~Chuang}, \bibfnamefont {N.~Gershenfeld}}\ and\ \bibinfo
  {author} {\bibfnamefont {D.~W.}\ \bibnamefont {Leung}},\ }\href@noop {}
  {\bibfield  {journal} {\bibinfo  {journal} {Proc. R. Soc. Lond.,Ser A}\
  }\textbf {\bibinfo {volume} {454}},\ \bibinfo {pages} {447} (\bibinfo {year}
  {1998})}\BibitemShut {NoStop}%
\bibitem [{\citenamefont {Childs}\ \emph {et~al.}(2001)\citenamefont {Childs},
  \citenamefont {Chuang},\ and\ \citenamefont {Leung}}]{ChuangPRA2001}%
  \BibitemOpen
  \bibfield  {author} {\bibinfo {author} {\bibfnamefont {A.~M.}\ \bibnamefont
  {Childs}}, \bibinfo {author} {\bibfnamefont {I.~L.}\ \bibnamefont {Chuang}},
  \ and\ \bibinfo {author} {\bibfnamefont {D.~W.}\ \bibnamefont {Leung}},\
  }\href {\doibase 10.1103/PhysRevA.64.012314} {\bibfield  {journal} {\bibinfo
  {journal} {Phys. Rev. A}\ }\textbf {\bibinfo {volume} {64}},\ \bibinfo
  {pages} {012314} (\bibinfo {year} {2001})}\BibitemShut {NoStop}%
\bibitem [{\citenamefont {Weinstein}\ \emph {et~al.}(2004)\citenamefont
  {Weinstein}, \citenamefont {Havel}, \citenamefont {Emerson}, \citenamefont
  {Boulant}, \citenamefont {Saraceno}, \citenamefont {Lloyd},\ and\
  \citenamefont {Cory}}]{QPTofQFT}%
  \BibitemOpen
  \bibfield  {author} {\bibinfo {author} {\bibfnamefont {Y.~S.}\ \bibnamefont
  {Weinstein}}, \bibinfo {author} {\bibfnamefont {T.~F.}\ \bibnamefont
  {Havel}}, \bibinfo {author} {\bibfnamefont {J.}~\bibnamefont {Emerson}},
  \bibinfo {author} {\bibfnamefont {N.}~\bibnamefont {Boulant}}, \bibinfo
  {author} {\bibfnamefont {M.}~\bibnamefont {Saraceno}}, \bibinfo {author}
  {\bibfnamefont {S.}~\bibnamefont {Lloyd}}, \ and\ \bibinfo {author}
  {\bibfnamefont {D.~G.}\ \bibnamefont {Cory}},\ }\href {\doibase
  http://dx.doi.org/10.1063/1.1785151} {\bibfield  {journal} {\bibinfo
  {journal} {The Journal of Chemical Physics}\ }\textbf {\bibinfo {volume}
  {121}},\ \bibinfo {pages} {6117} (\bibinfo {year} {2004})}\BibitemShut
  {NoStop}%
\bibitem [{\citenamefont {De~Martini}\ \emph {et~al.}(2003)\citenamefont
  {De~Martini}, \citenamefont {Mazzei}, \citenamefont {Ricci},\ and\
  \citenamefont {D'Ariano}}]{EAPT1Exp}%
  \BibitemOpen
  \bibfield  {author} {\bibinfo {author} {\bibfnamefont {F.}~\bibnamefont
  {De~Martini}}, \bibinfo {author} {\bibfnamefont {A.}~\bibnamefont {Mazzei}},
  \bibinfo {author} {\bibfnamefont {M.}~\bibnamefont {Ricci}}, \ and\ \bibinfo
  {author} {\bibfnamefont {G.~M.}\ \bibnamefont {D'Ariano}},\ }\href {\doibase
  10.1103/PhysRevA.67.062307} {\bibfield  {journal} {\bibinfo  {journal} {Phys.
  Rev. A}\ }\textbf {\bibinfo {volume} {67}},\ \bibinfo {pages} {062307}
  (\bibinfo {year} {2003})}\BibitemShut {NoStop}%
\bibitem [{\citenamefont {Altepeter}\ \emph {et~al.}(2003)\citenamefont
  {Altepeter}, \citenamefont {Branning}, \citenamefont {Jeffrey}, \citenamefont
  {Wei}, \citenamefont {Kwiat}, \citenamefont {Thew}, \citenamefont {O'Brien},
  \citenamefont {Nielsen},\ and\ \citenamefont {White}}]{altepeter}%
  \BibitemOpen
  \bibfield  {author} {\bibinfo {author} {\bibfnamefont {J.~B.}\ \bibnamefont
  {Altepeter}}, \bibinfo {author} {\bibfnamefont {D.}~\bibnamefont {Branning}},
  \bibinfo {author} {\bibfnamefont {E.}~\bibnamefont {Jeffrey}}, \bibinfo
  {author} {\bibfnamefont {T.~C.}\ \bibnamefont {Wei}}, \bibinfo {author}
  {\bibfnamefont {P.~G.}\ \bibnamefont {Kwiat}}, \bibinfo {author}
  {\bibfnamefont {R.~T.}\ \bibnamefont {Thew}}, \bibinfo {author}
  {\bibfnamefont {J.~L.}\ \bibnamefont {O'Brien}}, \bibinfo {author}
  {\bibfnamefont {M.~A.}\ \bibnamefont {Nielsen}}, \ and\ \bibinfo {author}
  {\bibfnamefont {A.~G.}\ \bibnamefont {White}},\ }\href {\doibase
  10.1103/PhysRevLett.90.193601} {\bibfield  {journal} {\bibinfo  {journal}
  {Phys. Rev. Lett.}\ }\textbf {\bibinfo {volume} {90}},\ \bibinfo {pages}
  {193601} (\bibinfo {year} {2003})}\BibitemShut {NoStop}%
\bibitem [{\citenamefont {O'Brien}\ \emph {et~al.}(2004)\citenamefont
  {O'Brien}, \citenamefont {Pryde}, \citenamefont {Gilchrist}, \citenamefont
  {James}, \citenamefont {Langford}, \citenamefont {Ralph},\ and\ \citenamefont
  {White}}]{obrian}%
  \BibitemOpen
  \bibfield  {author} {\bibinfo {author} {\bibfnamefont {J.~L.}\ \bibnamefont
  {O'Brien}}, \bibinfo {author} {\bibfnamefont {G.~J.}\ \bibnamefont {Pryde}},
  \bibinfo {author} {\bibfnamefont {A.}~\bibnamefont {Gilchrist}}, \bibinfo
  {author} {\bibfnamefont {D.~F.~V.}\ \bibnamefont {James}}, \bibinfo {author}
  {\bibfnamefont {N.~K.}\ \bibnamefont {Langford}}, \bibinfo {author}
  {\bibfnamefont {T.~C.}\ \bibnamefont {Ralph}}, \ and\ \bibinfo {author}
  {\bibfnamefont {A.~G.}\ \bibnamefont {White}},\ }\href {\doibase
  10.1103/PhysRevLett.93.080502} {\bibfield  {journal} {\bibinfo  {journal}
  {Phys. Rev. Lett.}\ }\textbf {\bibinfo {volume} {93}},\ \bibinfo {pages}
  {080502} (\bibinfo {year} {2004})}\BibitemShut {NoStop}%
\bibitem [{\citenamefont {Mitchell}\ \emph {et~al.}(2003)\citenamefont
  {Mitchell}, \citenamefont {Ellenor}, \citenamefont {Schneider},\ and\
  \citenamefont {Steinberg}}]{bellstatefilter}%
  \BibitemOpen
  \bibfield  {author} {\bibinfo {author} {\bibfnamefont {M.~W.}\ \bibnamefont
  {Mitchell}}, \bibinfo {author} {\bibfnamefont {C.~W.}\ \bibnamefont
  {Ellenor}}, \bibinfo {author} {\bibfnamefont {S.}~\bibnamefont {Schneider}},
  \ and\ \bibinfo {author} {\bibfnamefont {A.~M.}\ \bibnamefont {Steinberg}},\
  }\href {\doibase 10.1103/PhysRevLett.91.120402} {\bibfield  {journal}
  {\bibinfo  {journal} {Phys. Rev. Lett.}\ }\textbf {\bibinfo {volume} {91}},\
  \bibinfo {pages} {120402} (\bibinfo {year} {2003})}\BibitemShut {NoStop}%
\bibitem [{\citenamefont {Riebe}\ \emph {et~al.}(2006)\citenamefont {Riebe},
  \citenamefont {Kim}, \citenamefont {Schindler}, \citenamefont {Monz},
  \citenamefont {Schmidt}, \citenamefont {K\"orber}, \citenamefont {H\"ansel},
  \citenamefont {H\"affner}, \citenamefont {Roos},\ and\ \citenamefont
  {Blatt}}]{qptITprl2006}%
  \BibitemOpen
  \bibfield  {author} {\bibinfo {author} {\bibfnamefont {M.}~\bibnamefont
  {Riebe}}, \bibinfo {author} {\bibfnamefont {K.}~\bibnamefont {Kim}}, \bibinfo
  {author} {\bibfnamefont {P.}~\bibnamefont {Schindler}}, \bibinfo {author}
  {\bibfnamefont {T.}~\bibnamefont {Monz}}, \bibinfo {author} {\bibfnamefont
  {P.~O.}\ \bibnamefont {Schmidt}}, \bibinfo {author} {\bibfnamefont {T.~K.}\
  \bibnamefont {K\"orber}}, \bibinfo {author} {\bibfnamefont {W.}~\bibnamefont
  {H\"ansel}}, \bibinfo {author} {\bibfnamefont {H.}~\bibnamefont {H\"affner}},
  \bibinfo {author} {\bibfnamefont {C.~F.}\ \bibnamefont {Roos}}, \ and\
  \bibinfo {author} {\bibfnamefont {R.}~\bibnamefont {Blatt}},\ }\href
  {\doibase 10.1103/PhysRevLett.97.220407} {\bibfield  {journal} {\bibinfo
  {journal} {Phys. Rev. Lett.}\ }\textbf {\bibinfo {volume} {97}},\ \bibinfo
  {pages} {220407} (\bibinfo {year} {2006})}\BibitemShut {NoStop}%
\bibitem [{\citenamefont {Hanneke}\ \emph {et~al.}(2010)\citenamefont
  {Hanneke}, \citenamefont {Home}, \citenamefont {Jost}, \citenamefont {Amini},
  \citenamefont {Leibfried},\ and\ \citenamefont
  {Wineland}}]{QPT_IonTrapNature2010}%
  \BibitemOpen
  \bibfield  {author} {\bibinfo {author} {\bibfnamefont {D.}~\bibnamefont
  {Hanneke}}, \bibinfo {author} {\bibfnamefont {J.~P.}\ \bibnamefont {Home}},
  \bibinfo {author} {\bibfnamefont {J.~D.}\ \bibnamefont {Jost}}, \bibinfo
  {author} {\bibfnamefont {J.~M.}\ \bibnamefont {Amini}}, \bibinfo {author}
  {\bibfnamefont {D.}~\bibnamefont {Leibfried}}, \ and\ \bibinfo {author}
  {\bibfnamefont {D.~J.}\ \bibnamefont {Wineland}},\ }\href {\doibase
  10.1038/nphys1453} {\bibfield  {journal} {\bibinfo  {journal} {Nature Phys.}\
  }\textbf {\bibinfo {volume} {6}},\ \bibinfo {pages} {13} (\bibinfo {year}
  {2010})}\BibitemShut {NoStop}%
\bibitem [{\citenamefont {Neeley}\ \emph {et~al.}(2008)\citenamefont {Neeley},
  \citenamefont {Ansmann}, \citenamefont {Bialczak}, \citenamefont {Hofheinz},
  \citenamefont {Katz}, \citenamefont {Lucero}, \citenamefont {O/'Connell},
  \citenamefont {Wang}, \citenamefont {Cleland},\ and\ \citenamefont
  {Martinis}}]{QPT_SQUID}%
  \BibitemOpen
  \bibfield  {author} {\bibinfo {author} {\bibfnamefont {M.}~\bibnamefont
  {Neeley}}, \bibinfo {author} {\bibfnamefont {M.}~\bibnamefont {Ansmann}},
  \bibinfo {author} {\bibfnamefont {R.~C.}\ \bibnamefont {Bialczak}}, \bibinfo
  {author} {\bibfnamefont {M.}~\bibnamefont {Hofheinz}}, \bibinfo {author}
  {\bibfnamefont {N.}~\bibnamefont {Katz}}, \bibinfo {author} {\bibfnamefont
  {E.}~\bibnamefont {Lucero}}, \bibinfo {author} {\bibfnamefont
  {A.}~\bibnamefont {O/'Connell}}, \bibinfo {author} {\bibfnamefont
  {H.}~\bibnamefont {Wang}}, \bibinfo {author} {\bibfnamefont {A.~N.}\
  \bibnamefont {Cleland}}, \ and\ \bibinfo {author} {\bibfnamefont {J.~M.}\
  \bibnamefont {Martinis}},\ }\href {\doibase 10.1038/nphys972} {\bibfield
  {journal} {\bibinfo  {journal} {Nature Phys.}\ }\textbf {\bibinfo {volume}
  {4}},\ \bibinfo {pages} {523} (\bibinfo {year} {2008})}\BibitemShut {NoStop}%
\bibitem [{\citenamefont {Chow}\ \emph {et~al.}(2009)\citenamefont {Chow},
  \citenamefont {Gambetta}, \citenamefont {Tornberg}, \citenamefont {Koch},
  \citenamefont {Bishop}, \citenamefont {Houck}, \citenamefont {Johnson},
  \citenamefont {Frunzio}, \citenamefont {Girvin},\ and\ \citenamefont
  {Schoelkopf}}]{SQUID2009}%
  \BibitemOpen
  \bibfield  {author} {\bibinfo {author} {\bibfnamefont {J.~M.}\ \bibnamefont
  {Chow}}, \bibinfo {author} {\bibfnamefont {J.~M.}\ \bibnamefont {Gambetta}},
  \bibinfo {author} {\bibfnamefont {L.}~\bibnamefont {Tornberg}}, \bibinfo
  {author} {\bibfnamefont {J.}~\bibnamefont {Koch}}, \bibinfo {author}
  {\bibfnamefont {L.~S.}\ \bibnamefont {Bishop}}, \bibinfo {author}
  {\bibfnamefont {A.~A.}\ \bibnamefont {Houck}}, \bibinfo {author}
  {\bibfnamefont {B.~R.}\ \bibnamefont {Johnson}}, \bibinfo {author}
  {\bibfnamefont {L.}~\bibnamefont {Frunzio}}, \bibinfo {author} {\bibfnamefont
  {S.~M.}\ \bibnamefont {Girvin}}, \ and\ \bibinfo {author} {\bibfnamefont
  {R.~J.}\ \bibnamefont {Schoelkopf}},\ }\href {\doibase
  10.1103/PhysRevLett.102.090502} {\bibfield  {journal} {\bibinfo  {journal}
  {Phys. Rev. Lett.}\ }\textbf {\bibinfo {volume} {102}},\ \bibinfo {pages}
  {090502} (\bibinfo {year} {2009})}\BibitemShut {NoStop}%
\bibitem [{\citenamefont {Bialczak}\ \emph {et~al.}(2010)\citenamefont
  {Bialczak}, \citenamefont {Ansmann}, \citenamefont {Hofheinz}, \citenamefont
  {Lucero}, \citenamefont {Neeley}, \citenamefont {O/'Connell}, \citenamefont
  {Sank}, \citenamefont {Wang}, \citenamefont {Wenner}, \citenamefont
  {Steffen},\ and\ \citenamefont {Cleland}}]{MartiniSQUID2010}%
  \BibitemOpen
  \bibfield  {author} {\bibinfo {author} {\bibfnamefont {R.~C.}\ \bibnamefont
  {Bialczak}}, \bibinfo {author} {\bibfnamefont {M.}~\bibnamefont {Ansmann}},
  \bibinfo {author} {\bibfnamefont {M.}~\bibnamefont {Hofheinz}}, \bibinfo
  {author} {\bibfnamefont {E.}~\bibnamefont {Lucero}}, \bibinfo {author}
  {\bibfnamefont {M.}~\bibnamefont {Neeley}}, \bibinfo {author} {\bibfnamefont
  {A.~D.}\ \bibnamefont {O/'Connell}}, \bibinfo {author} {\bibfnamefont
  {D.}~\bibnamefont {Sank}}, \bibinfo {author} {\bibfnamefont {H.}~\bibnamefont
  {Wang}}, \bibinfo {author} {\bibfnamefont {J.}~\bibnamefont {Wenner}},
  \bibinfo {author} {\bibfnamefont {M.}~\bibnamefont {Steffen}}, \ and\
  \bibinfo {author} {\bibfnamefont {J.~M.}\ \bibnamefont {Cleland},
  \bibfnamefont {A.~N.and~Martinis}},\ }\href {\doibase 10.1038/nphys1639}
  {\bibfield  {journal} {\bibinfo  {journal} {Nature Phys.}\ }\textbf {\bibinfo
  {volume} {6}},\ \bibinfo {pages} {409} (\bibinfo {year} {2010})}\BibitemShut
  {NoStop}%
\bibitem [{\citenamefont {Yamamoto}\ \emph {et~al.}(2010)\citenamefont
  {Yamamoto}, \citenamefont {Neeley}, \citenamefont {Lucero}, \citenamefont
  {Bialczak}, \citenamefont {Kelly}, \citenamefont {Lenander}, \citenamefont
  {Mariantoni}, \citenamefont {O'Connell}, \citenamefont {Sank}, \citenamefont
  {Wang}, \citenamefont {Weides}, \citenamefont {Wenner}, \citenamefont {Yin},
  \citenamefont {Cleland},\ and\ \citenamefont {Martinis}}]{QPT2spSQUID}%
  \BibitemOpen
  \bibfield  {author} {\bibinfo {author} {\bibfnamefont {T.}~\bibnamefont
  {Yamamoto}}, \bibinfo {author} {\bibfnamefont {M.}~\bibnamefont {Neeley}},
  \bibinfo {author} {\bibfnamefont {E.}~\bibnamefont {Lucero}}, \bibinfo
  {author} {\bibfnamefont {R.~C.}\ \bibnamefont {Bialczak}}, \bibinfo {author}
  {\bibfnamefont {J.}~\bibnamefont {Kelly}}, \bibinfo {author} {\bibfnamefont
  {M.}~\bibnamefont {Lenander}}, \bibinfo {author} {\bibfnamefont
  {M.}~\bibnamefont {Mariantoni}}, \bibinfo {author} {\bibfnamefont {A.~D.}\
  \bibnamefont {O'Connell}}, \bibinfo {author} {\bibfnamefont {D.}~\bibnamefont
  {Sank}}, \bibinfo {author} {\bibfnamefont {H.}~\bibnamefont {Wang}}, \bibinfo
  {author} {\bibfnamefont {M.}~\bibnamefont {Weides}}, \bibinfo {author}
  {\bibfnamefont {J.}~\bibnamefont {Wenner}}, \bibinfo {author} {\bibfnamefont
  {Y.}~\bibnamefont {Yin}}, \bibinfo {author} {\bibfnamefont {A.~N.}\
  \bibnamefont {Cleland}}, \ and\ \bibinfo {author} {\bibfnamefont {J.~M.}\
  \bibnamefont {Martinis}},\ }\href {\doibase 10.1103/PhysRevB.82.184515}
  {\bibfield  {journal} {\bibinfo  {journal} {Phys. Rev. B}\ }\textbf {\bibinfo
  {volume} {82}},\ \bibinfo {pages} {184515} (\bibinfo {year}
  {2010})}\BibitemShut {NoStop}%
\bibitem [{\citenamefont {Chow}\ \emph {et~al.}(2011)\citenamefont {Chow},
  \citenamefont {C\'orcoles}, \citenamefont {Gambetta}, \citenamefont
  {Rigetti}, \citenamefont {Johnson}, \citenamefont {Smolin}, \citenamefont
  {Rozen}, \citenamefont {Keefe}, \citenamefont {Rothwell}, \citenamefont
  {Ketchen},\ and\ \citenamefont {Steffen}}]{QPT_SQUIDChow2011}%
  \BibitemOpen
  \bibfield  {author} {\bibinfo {author} {\bibfnamefont {J.~M.}\ \bibnamefont
  {Chow}}, \bibinfo {author} {\bibfnamefont {A.~D.}\ \bibnamefont
  {C\'orcoles}}, \bibinfo {author} {\bibfnamefont {J.~M.}\ \bibnamefont
  {Gambetta}}, \bibinfo {author} {\bibfnamefont {C.}~\bibnamefont {Rigetti}},
  \bibinfo {author} {\bibfnamefont {B.~R.}\ \bibnamefont {Johnson}}, \bibinfo
  {author} {\bibfnamefont {J.~A.}\ \bibnamefont {Smolin}}, \bibinfo {author}
  {\bibfnamefont {J.~R.}\ \bibnamefont {Rozen}}, \bibinfo {author}
  {\bibfnamefont {G.~A.}\ \bibnamefont {Keefe}}, \bibinfo {author}
  {\bibfnamefont {M.~B.}\ \bibnamefont {Rothwell}}, \bibinfo {author}
  {\bibfnamefont {M.~B.}\ \bibnamefont {Ketchen}}, \ and\ \bibinfo {author}
  {\bibfnamefont {M.}~\bibnamefont {Steffen}},\ }\href {\doibase
  10.1103/PhysRevLett.107.080502} {\bibfield  {journal} {\bibinfo  {journal}
  {Phys. Rev. Lett.}\ }\textbf {\bibinfo {volume} {107}},\ \bibinfo {pages}
  {080502} (\bibinfo {year} {2011})}\BibitemShut {NoStop}%
\bibitem [{\citenamefont {Dewes}\ \emph {et~al.}(2012)\citenamefont {Dewes},
  \citenamefont {Ong}, \citenamefont {Schmitt}, \citenamefont {Lauro},
  \citenamefont {Boulant}, \citenamefont {Bertet}, \citenamefont {Vion},\ and\
  \citenamefont {Esteve}}]{SQUID_Dewes2012}%
  \BibitemOpen
  \bibfield  {author} {\bibinfo {author} {\bibfnamefont {A.}~\bibnamefont
  {Dewes}}, \bibinfo {author} {\bibfnamefont {F.~R.}\ \bibnamefont {Ong}},
  \bibinfo {author} {\bibfnamefont {V.}~\bibnamefont {Schmitt}}, \bibinfo
  {author} {\bibfnamefont {R.}~\bibnamefont {Lauro}}, \bibinfo {author}
  {\bibfnamefont {N.}~\bibnamefont {Boulant}}, \bibinfo {author} {\bibfnamefont
  {P.}~\bibnamefont {Bertet}}, \bibinfo {author} {\bibfnamefont
  {D.}~\bibnamefont {Vion}}, \ and\ \bibinfo {author} {\bibfnamefont
  {D.}~\bibnamefont {Esteve}},\ }\href {\doibase
  10.1103/PhysRevLett.108.057002} {\bibfield  {journal} {\bibinfo  {journal}
  {Phys. Rev. Lett.}\ }\textbf {\bibinfo {volume} {108}},\ \bibinfo {pages}
  {057002} (\bibinfo {year} {2012})}\BibitemShut {NoStop}%
\bibitem [{\citenamefont {Zhang}\ \emph {et~al.}(2014)\citenamefont {Zhang},
  \citenamefont {Souza}, \citenamefont {Brandao},\ and\ \citenamefont
  {Suter}}]{suterprotectedgate}%
  \BibitemOpen
  \bibfield  {author} {\bibinfo {author} {\bibfnamefont {J.}~\bibnamefont
  {Zhang}}, \bibinfo {author} {\bibfnamefont {A.~M.}\ \bibnamefont {Souza}},
  \bibinfo {author} {\bibfnamefont {F.~D.}\ \bibnamefont {Brandao}}, \ and\
  \bibinfo {author} {\bibfnamefont {D.}~\bibnamefont {Suter}},\ }\href
  {\doibase 10.1103/PhysRevLett.112.050502} {\bibfield  {journal} {\bibinfo
  {journal} {Phys. Rev. Lett.}\ }\textbf {\bibinfo {volume} {112}},\ \bibinfo
  {pages} {050502} (\bibinfo {year} {2014})}\BibitemShut {NoStop}%
\bibitem [{\citenamefont {Shabani}\ \emph {et~al.}(2011)\citenamefont
  {Shabani}, \citenamefont {Kosut}, \citenamefont {Mohseni}, \citenamefont
  {Rabitz}, \citenamefont {Broome}, \citenamefont {Almeida}, \citenamefont
  {Fedrizzi},\ and\ \citenamefont {White}}]{simplifiedQPT1}%
  \BibitemOpen
  \bibfield  {author} {\bibinfo {author} {\bibfnamefont {A.}~\bibnamefont
  {Shabani}}, \bibinfo {author} {\bibfnamefont {R.~L.}\ \bibnamefont {Kosut}},
  \bibinfo {author} {\bibfnamefont {M.}~\bibnamefont {Mohseni}}, \bibinfo
  {author} {\bibfnamefont {H.}~\bibnamefont {Rabitz}}, \bibinfo {author}
  {\bibfnamefont {M.~A.}\ \bibnamefont {Broome}}, \bibinfo {author}
  {\bibfnamefont {M.~P.}\ \bibnamefont {Almeida}}, \bibinfo {author}
  {\bibfnamefont {A.}~\bibnamefont {Fedrizzi}}, \ and\ \bibinfo {author}
  {\bibfnamefont {A.~G.}\ \bibnamefont {White}},\ }\href {\doibase
  10.1103/PhysRevLett.106.100401} {\bibfield  {journal} {\bibinfo  {journal}
  {Phys. Rev. Lett.}\ }\textbf {\bibinfo {volume} {106}},\ \bibinfo {pages}
  {100401} (\bibinfo {year} {2011})}\BibitemShut {NoStop}%
\bibitem [{\citenamefont {Wu}\ \emph {et~al.}(2013)\citenamefont {Wu},
  \citenamefont {Li}, \citenamefont {Zheng}, \citenamefont {Peng},\ and\
  \citenamefont {Feng}}]{simplifiedQPT2}%
  \BibitemOpen
  \bibfield  {author} {\bibinfo {author} {\bibfnamefont {Z.}~\bibnamefont
  {Wu}}, \bibinfo {author} {\bibfnamefont {S.}~\bibnamefont {Li}}, \bibinfo
  {author} {\bibfnamefont {W.}~\bibnamefont {Zheng}}, \bibinfo {author}
  {\bibfnamefont {X.}~\bibnamefont {Peng}}, \ and\ \bibinfo {author}
  {\bibfnamefont {M.}~\bibnamefont {Feng}},\ }\href {\doibase
  http://dx.doi.org/10.1063/1.4774119} {\bibfield  {journal} {\bibinfo
  {journal} {The Journal of Chemical Physics}\ }\textbf {\bibinfo {volume}
  {138}},\ \bibinfo {pages} {024318} (\bibinfo {year} {2013})}\BibitemShut
  {NoStop}%
\bibitem [{\citenamefont {Mazzei}\ \emph {et~al.}(2003)\citenamefont {Mazzei},
  \citenamefont {Ricci}, \citenamefont {De~Martini},\ and\ \citenamefont
  {D'Ariano}}]{mazzei2003pauli}%
  \BibitemOpen
  \bibfield  {author} {\bibinfo {author} {\bibfnamefont {A.}~\bibnamefont
  {Mazzei}}, \bibinfo {author} {\bibfnamefont {M.}~\bibnamefont {Ricci}},
  \bibinfo {author} {\bibfnamefont {F.}~\bibnamefont {De~Martini}}, \ and\
  \bibinfo {author} {\bibfnamefont {G.}~\bibnamefont {D'Ariano}},\ }\href@noop
  {} {\bibfield  {journal} {\bibinfo  {journal} {Fortschritte der Physik}\
  }\textbf {\bibinfo {volume} {51}},\ \bibinfo {pages} {342} (\bibinfo {year}
  {2003})}\BibitemShut {NoStop}%
\bibitem [{\citenamefont {D'Ariano}\ and\ \citenamefont
  {Lo~Presti}(2003)}]{PRLAriano2003}%
  \BibitemOpen
  \bibfield  {author} {\bibinfo {author} {\bibfnamefont {G.~M.}\ \bibnamefont
  {D'Ariano}}\ and\ \bibinfo {author} {\bibfnamefont {P.}~\bibnamefont
  {Lo~Presti}},\ }\href {\doibase 10.1103/PhysRevLett.91.047902} {\bibfield
  {journal} {\bibinfo  {journal} {Phys. Rev. Lett.}\ }\textbf {\bibinfo
  {volume} {91}},\ \bibinfo {pages} {047902} (\bibinfo {year}
  {2003})}\BibitemShut {NoStop}%
\bibitem [{\citenamefont {Allahverdyan}\ \emph {et~al.}(2004)\citenamefont
  {Allahverdyan}, \citenamefont {Balian},\ and\ \citenamefont
  {Nieuwenhuizen}}]{Allahverdyan}%
  \BibitemOpen
  \bibfield  {author} {\bibinfo {author} {\bibfnamefont {A.~E.}\ \bibnamefont
  {Allahverdyan}}, \bibinfo {author} {\bibfnamefont {R.}~\bibnamefont
  {Balian}}, \ and\ \bibinfo {author} {\bibfnamefont {T.~M.}\ \bibnamefont
  {Nieuwenhuizen}},\ }\href {\doibase 10.1103/PhysRevLett.92.120402} {\bibfield
   {journal} {\bibinfo  {journal} {Phys. Rev. Lett.}\ }\textbf {\bibinfo
  {volume} {92}},\ \bibinfo {pages} {120402} (\bibinfo {year}
  {2004})}\BibitemShut {NoStop}%
\bibitem [{\citenamefont {Peng}\ \emph {et~al.}(2007)\citenamefont {Peng},
  \citenamefont {Du},\ and\ \citenamefont {Suter}}]{Suteraaqst}%
  \BibitemOpen
  \bibfield  {author} {\bibinfo {author} {\bibfnamefont {X.}~\bibnamefont
  {Peng}}, \bibinfo {author} {\bibfnamefont {J.}~\bibnamefont {Du}}, \ and\
  \bibinfo {author} {\bibfnamefont {D.}~\bibnamefont {Suter}},\ }\href
  {\doibase 10.1103/PhysRevA.76.042117} {\bibfield  {journal} {\bibinfo
  {journal} {Phys. Rev. A}\ }\textbf {\bibinfo {volume} {76}},\ \bibinfo
  {pages} {042117} (\bibinfo {year} {2007})}\BibitemShut {NoStop}%
\bibitem [{\citenamefont {Yu}\ \emph {et~al.}(2011)\citenamefont {Yu},
  \citenamefont {Wen}, \citenamefont {Li},\ and\ \citenamefont {Peng}}]{peng}%
  \BibitemOpen
  \bibfield  {author} {\bibinfo {author} {\bibfnamefont {Y.}~\bibnamefont
  {Yu}}, \bibinfo {author} {\bibfnamefont {H.}~\bibnamefont {Wen}}, \bibinfo
  {author} {\bibfnamefont {H.}~\bibnamefont {Li}}, \ and\ \bibinfo {author}
  {\bibfnamefont {X.}~\bibnamefont {Peng}},\ }\href {\doibase
  10.1103/PhysRevA.83.032318} {\bibfield  {journal} {\bibinfo  {journal} {Phys.
  Rev. A}\ }\textbf {\bibinfo {volume} {83}},\ \bibinfo {pages} {032318}
  (\bibinfo {year} {2011})}\BibitemShut {NoStop}%
\bibitem [{\citenamefont {Shukla}\ \emph {et~al.}(2013)\citenamefont {Shukla},
  \citenamefont {Rao},\ and\ \citenamefont {Mahesh}}]{abhishek}%
  \BibitemOpen
  \bibfield  {author} {\bibinfo {author} {\bibfnamefont {A.}~\bibnamefont
  {Shukla}}, \bibinfo {author} {\bibfnamefont {K.~R.~K.}\ \bibnamefont {Rao}},
  \ and\ \bibinfo {author} {\bibfnamefont {T.~S.}\ \bibnamefont {Mahesh}},\
  }\href {\doibase 10.1103/PhysRevA.87.062317} {\bibfield  {journal} {\bibinfo
  {journal} {Phys. Rev. A}\ }\textbf {\bibinfo {volume} {87}},\ \bibinfo
  {pages} {062317} (\bibinfo {year} {2013})}\BibitemShut {NoStop}%
\bibitem [{\citenamefont {Nielsen}\ and\ \citenamefont
  {Chuang}(2010)}]{chuangbook}%
  \BibitemOpen
  \bibfield  {author} {\bibinfo {author} {\bibnamefont {Nielsen}}\ and\
  \bibinfo {author} {\bibfnamefont {I.~L.}\ \bibnamefont {Chuang}},\
  }\href@noop {} {\emph {\bibinfo {title} {Quantum computation and quantum
  information}}}\ (\bibinfo  {publisher} {Cambridge university press},\
  \bibinfo {year} {2010})\BibitemShut {NoStop}%
\bibitem [{\citenamefont {Cavanagh}\ \emph {et~al.}(1995)\citenamefont
  {Cavanagh}, \citenamefont {Fairbrother}, \citenamefont {Palmer~III},\ and\
  \citenamefont {Skelton}}]{cavanagh}%
  \BibitemOpen
  \bibfield  {author} {\bibinfo {author} {\bibfnamefont {J.}~\bibnamefont
  {Cavanagh}}, \bibinfo {author} {\bibfnamefont {W.~J.}\ \bibnamefont
  {Fairbrother}}, \bibinfo {author} {\bibfnamefont {A.~G.}\ \bibnamefont
  {Palmer~III}}, \ and\ \bibinfo {author} {\bibfnamefont {N.~J.}\ \bibnamefont
  {Skelton}},\ }\href@noop {} {\emph {\bibinfo {title} {Protein NMR
  spectroscopy: principles and practice}}}\ (\bibinfo  {publisher} {Academic
  Press},\ \bibinfo {year} {1995})\BibitemShut {NoStop}%
\bibitem [{\citenamefont {Khaneja}\ \emph {et~al.}(2005)\citenamefont
  {Khaneja}, \citenamefont {Reiss}, \citenamefont {Kehlet}, \citenamefont
  {Schulte-Herbr{\"u}ggen},\ and\ \citenamefont {Glaser}}]{khaneja2005optimal}%
  \BibitemOpen
  \bibfield  {author} {\bibinfo {author} {\bibfnamefont {N.}~\bibnamefont
  {Khaneja}}, \bibinfo {author} {\bibfnamefont {T.}~\bibnamefont {Reiss}},
  \bibinfo {author} {\bibfnamefont {C.}~\bibnamefont {Kehlet}}, \bibinfo
  {author} {\bibfnamefont {T.}~\bibnamefont {Schulte-Herbr{\"u}ggen}}, \ and\
  \bibinfo {author} {\bibfnamefont {S.~J.}\ \bibnamefont {Glaser}},\
  }\href@noop {} {\bibfield  {journal} {\bibinfo  {journal} {Journal of
  Magnetic Resonance}\ }\textbf {\bibinfo {volume} {172}},\ \bibinfo {pages}
  {296} (\bibinfo {year} {2005})}\BibitemShut {NoStop}%
\bibitem [{\citenamefont {Cory}\ \emph {et~al.}(1997)\citenamefont {Cory},
  \citenamefont {Fahmy},\ and\ \citenamefont {Havel}}]{cory2000nmr}%
  \BibitemOpen
  \bibfield  {author} {\bibinfo {author} {\bibfnamefont {D.~G.}\ \bibnamefont
  {Cory}}, \bibinfo {author} {\bibfnamefont {A.~F.}\ \bibnamefont {Fahmy}}, \
  and\ \bibinfo {author} {\bibfnamefont {T.~F.}\ \bibnamefont {Havel}},\ }\href
  {http://www.pnas.org/content/94/5/1634.full} {\bibfield  {journal} {\bibinfo
  {journal} {Proc. Natl. Acad. Sci. USA}\ }\textbf {\bibinfo {volume} {94}},\
  \bibinfo {pages} {1634} (\bibinfo {year} {1997})}\BibitemShut {NoStop}%
\bibitem [{sup()}]{sup}%
  \BibitemOpen
  \href@noop {} {\bibinfo  {journal} {see supplementary material}\
  }\BibitemShut {NoStop}%
\bibitem [{\citenamefont {Bennett}\ \emph
  {et~al.}(1996{\natexlab{a}})\citenamefont {Bennett}, \citenamefont
  {Brassard}, \citenamefont {Popescu}, \citenamefont {Schumacher},
  \citenamefont {Smolin},\ and\ \citenamefont {Wootters}}]{benett1}%
  \BibitemOpen
\bibfield  {journal} {  }\bibfield  {author} {\bibinfo {author} {\bibfnamefont
  {C.~H.}\ \bibnamefont {Bennett}}, \bibinfo {author} {\bibfnamefont
  {G.}~\bibnamefont {Brassard}}, \bibinfo {author} {\bibfnamefont
  {S.}~\bibnamefont {Popescu}}, \bibinfo {author} {\bibfnamefont
  {B.}~\bibnamefont {Schumacher}}, \bibinfo {author} {\bibfnamefont {J.~A.}\
  \bibnamefont {Smolin}}, \ and\ \bibinfo {author} {\bibfnamefont {W.~K.}\
  \bibnamefont {Wootters}},\ }\href {\doibase 10.1088/1464-4266/7/10/021}
  {\bibfield  {journal} {\bibinfo  {journal} {Phys. Rev. Lett.}\ }\textbf
  {\bibinfo {volume} {76}},\ \bibinfo {pages} {722} (\bibinfo {year}
  {1996}{\natexlab{a}})}\BibitemShut {NoStop}%
\bibitem [{\citenamefont {Bennett}\ \emph
  {et~al.}(1996{\natexlab{b}})\citenamefont {Bennett}, \citenamefont
  {DiVincenzo}, \citenamefont {Smolin},\ and\ \citenamefont
  {Wootters}}]{benett2}%
  \BibitemOpen
  \bibfield  {author} {\bibinfo {author} {\bibfnamefont {C.~H.}\ \bibnamefont
  {Bennett}}, \bibinfo {author} {\bibfnamefont {D.~P.}\ \bibnamefont
  {DiVincenzo}}, \bibinfo {author} {\bibfnamefont {J.~A.}\ \bibnamefont
  {Smolin}}, \ and\ \bibinfo {author} {\bibfnamefont {W.~K.}\ \bibnamefont
  {Wootters}},\ }\href {\doibase 10.1103/PhysRevA.54.3824} {\bibfield
  {journal} {\bibinfo  {journal} {Phys. Rev. A}\ }\textbf {\bibinfo {volume}
  {54}},\ \bibinfo {pages} {3824} (\bibinfo {year}
  {1996}{\natexlab{b}})}\BibitemShut {NoStop}%
\bibitem [{\citenamefont {Emerson}\ \emph {et~al.}(2005)\citenamefont
  {Emerson}, \citenamefont {Alicki},\ and\ \citenamefont
  {Życzkowski}}]{emerson1}%
  \BibitemOpen
  \bibfield  {author} {\bibinfo {author} {\bibfnamefont {J.}~\bibnamefont
  {Emerson}}, \bibinfo {author} {\bibfnamefont {R.}~\bibnamefont {Alicki}}, \
  and\ \bibinfo {author} {\bibfnamefont {K.}~\bibnamefont {Życzkowski}},\
  }\href {\doibase 10.1103/PhysRevA.54.3824} {\bibfield  {journal} {\bibinfo
  {journal} {J. Opt. B: Quantum Semiclass. Opt.}\ }\textbf {\bibinfo {volume}
  {7}},\ \bibinfo {pages} {S347} (\bibinfo {year} {2005})}\BibitemShut
  {NoStop}%
\bibitem [{\citenamefont {Emerson}\ \emph {et~al.}(2007)\citenamefont
  {Emerson}, \citenamefont {Silva}, \citenamefont {Moussa}, \citenamefont
  {Ryan}, \citenamefont {Laforest}, \citenamefont {Baugh}, \citenamefont
  {Cory},\ and\ \citenamefont {Laflamme}}]{emerson2}%
  \BibitemOpen
  \bibfield  {author} {\bibinfo {author} {\bibfnamefont {J.}~\bibnamefont
  {Emerson}}, \bibinfo {author} {\bibfnamefont {M.}~\bibnamefont {Silva}},
  \bibinfo {author} {\bibfnamefont {O.}~\bibnamefont {Moussa}}, \bibinfo
  {author} {\bibfnamefont {C.}~\bibnamefont {Ryan}}, \bibinfo {author}
  {\bibfnamefont {M.}~\bibnamefont {Laforest}}, \bibinfo {author}
  {\bibfnamefont {J.}~\bibnamefont {Baugh}}, \bibinfo {author} {\bibfnamefont
  {D.~G.}\ \bibnamefont {Cory}}, \ and\ \bibinfo {author} {\bibfnamefont
  {R.}~\bibnamefont {Laflamme}},\ }\href {\doibase 10.1126/science.1145699}
  {\bibfield  {journal} {\bibinfo  {journal} {Science}\ }\textbf {\bibinfo
  {volume} {317}},\ \bibinfo {pages} {1893} (\bibinfo {year}
  {2007})}\BibitemShut {NoStop}%
\bibitem [{\citenamefont {Silva}\ \emph {et~al.}(2008)\citenamefont {Silva},
  \citenamefont {Magesan}, \citenamefont {Kribs},\ and\ \citenamefont
  {Emerson}}]{emerson3}%
  \BibitemOpen
  \bibfield  {author} {\bibinfo {author} {\bibfnamefont {M.}~\bibnamefont
  {Silva}}, \bibinfo {author} {\bibfnamefont {E.}~\bibnamefont {Magesan}},
  \bibinfo {author} {\bibfnamefont {D.~W.}\ \bibnamefont {Kribs}}, \ and\
  \bibinfo {author} {\bibfnamefont {J.}~\bibnamefont {Emerson}},\ }\href
  {\doibase 10.1103/PhysRevA.78.012347} {\bibfield  {journal} {\bibinfo
  {journal} {Phys. Rev. A}\ }\textbf {\bibinfo {volume} {78}},\ \bibinfo
  {pages} {012347} (\bibinfo {year} {2008})}\BibitemShut {NoStop}%
\bibitem [{\citenamefont {L\'opez}\ \emph {et~al.}(2010)\citenamefont
  {L\'opez}, \citenamefont {Bendersky}, \citenamefont {Paz},\ and\
  \citenamefont {Cory}}]{corytwirl}%
  \BibitemOpen
  \bibfield  {author} {\bibinfo {author} {\bibfnamefont {C.~C.}\ \bibnamefont
  {L\'opez}}, \bibinfo {author} {\bibfnamefont {A.}~\bibnamefont {Bendersky}},
  \bibinfo {author} {\bibfnamefont {J.~P.}\ \bibnamefont {Paz}}, \ and\
  \bibinfo {author} {\bibfnamefont {D.~G.}\ \bibnamefont {Cory}},\ }\href
  {\doibase 10.1103/PhysRevA.81.062113} {\bibfield  {journal} {\bibinfo
  {journal} {Phys. Rev. A}\ }\textbf {\bibinfo {volume} {81}},\ \bibinfo
  {pages} {062113} (\bibinfo {year} {2010})}\BibitemShut {NoStop}%
\bibitem [{\citenamefont {Dankert}\ \emph {et~al.}(2009)\citenamefont
  {Dankert}, \citenamefont {Cleve}, \citenamefont {Emerson},\ and\
  \citenamefont {Livine}}]{dankert2009}%
  \BibitemOpen
  \bibfield  {author} {\bibinfo {author} {\bibfnamefont {C.}~\bibnamefont
  {Dankert}}, \bibinfo {author} {\bibfnamefont {R.}~\bibnamefont {Cleve}},
  \bibinfo {author} {\bibfnamefont {J.}~\bibnamefont {Emerson}}, \ and\
  \bibinfo {author} {\bibfnamefont {E.}~\bibnamefont {Livine}},\ }\href
  {\doibase 10.1103/PhysRevA.80.012304} {\bibfield  {journal} {\bibinfo
  {journal} {Phys. Rev. A}\ }\textbf {\bibinfo {volume} {80}},\ \bibinfo
  {pages} {012304} (\bibinfo {year} {2009})}\BibitemShut {NoStop}%
\bibitem [{\citenamefont {Moussa}\ \emph {et~al.}(2012)\citenamefont {Moussa},
  \citenamefont {da~Silva}, \citenamefont {Ryan},\ and\ \citenamefont
  {Laflamme}}]{laflammeprl2012}%
  \BibitemOpen
  \bibfield  {author} {\bibinfo {author} {\bibfnamefont {O.}~\bibnamefont
  {Moussa}}, \bibinfo {author} {\bibfnamefont {M.~P.}\ \bibnamefont
  {da~Silva}}, \bibinfo {author} {\bibfnamefont {C.~A.}\ \bibnamefont {Ryan}},
  \ and\ \bibinfo {author} {\bibfnamefont {R.}~\bibnamefont {Laflamme}},\
  }\href {\doibase 10.1103/PhysRevLett.109.070504} {\bibfield  {journal}
  {\bibinfo  {journal} {Phys. Rev. Lett.}\ }\textbf {\bibinfo {volume} {109}},\
  \bibinfo {pages} {070504} (\bibinfo {year} {2012})}\BibitemShut {NoStop}%
\bibitem [{\citenamefont {Anwar}\ \emph {et~al.}(2005)\citenamefont {Anwar},
  \citenamefont {Xiao}, \citenamefont {Short}, \citenamefont {Jones},
  \citenamefont {Blazina}, \citenamefont {Duckett},\ and\ \citenamefont
  {Carteret}}]{anwar}%
  \BibitemOpen
  \bibfield  {author} {\bibinfo {author} {\bibfnamefont {M.}~\bibnamefont
  {Anwar}}, \bibinfo {author} {\bibfnamefont {L.}~\bibnamefont {Xiao}},
  \bibinfo {author} {\bibfnamefont {A.}~\bibnamefont {Short}}, \bibinfo
  {author} {\bibfnamefont {J.}~\bibnamefont {Jones}}, \bibinfo {author}
  {\bibfnamefont {D.}~\bibnamefont {Blazina}}, \bibinfo {author} {\bibfnamefont
  {S.}~\bibnamefont {Duckett}}, \ and\ \bibinfo {author} {\bibfnamefont
  {H.}~\bibnamefont {Carteret}},\ }\href@noop {} {\bibfield  {journal}
  {\bibinfo  {journal} {Physical Review A}\ }\textbf {\bibinfo {volume} {71}},\
  \bibinfo {pages} {032327} (\bibinfo {year} {2005})}\BibitemShut {NoStop}%
\end{thebibliography}%
\end{document}


\begin{center}
{\large Supplementary materiel for \\ 
\vspace{0.2cm}
{\bf Single-Shot Quantum Process Tomography}}\\
\vspace{0.2cm}
{Abhishek Shukla and T. S. Mahesh}\\
{Department of Physics and NMR Research Center,\\
Indian Institute of Science Education and Research, Pune 411008, India}
\end{center}

Quantum process tomography (QPT) is a procedure for experimental characterization of a quantum process.
In a basis of fixed set of operators $\{ \tilde{E}_m \}$, a general process can be expressed in
$\chi$ matrix representation.  The $\chi$ matrix can be calculated using equation
\begin{eqnarray}
\beta \chi = \lambda,
\label{chi}
\end{eqnarray}

As explained in the text, given set $\{ \tilde{E}_m \}$ and linearly independent basis elements $\{ \rho_{j} \}$,
matrix $\beta$ can be calculated by 
 \begin{eqnarray}
 \tilde{E}_m\rho_j\tilde{E}_n^\dagger = \sum_k \beta_{jk}^{mn}\rho_k,
 \label{beta}
 \end{eqnarray}
 where $\beta_{jk}^{mn}$ is the element of matrix $\beta$.
 While elements of matrix $\lambda$ were extracted directly from 
 experimental output density matrix (see Eq. 8 in main text).
In our choice of ${\mathrm \{ \tilde{E}_m}\} = \{{\mathrm I}, {\mathrm X}, {\mathrm -iY}, {\mathrm Z}\}$
and  $\rho_{j}$ =  $\{\outpr{0}{0},\outpr{0}{1},\outpr{1}{0},\outpr{1}{1}\}$.
the $\beta$ matrix is following.

\begin{table}[h]
\caption{\Large $\beta$ matrix}
\vspace{0.7cm}
$
\begin{array}{|c|cccc|cccc|cccc|cccc|cccc|cccc|cccc|}

\hline
 {\{m,n\}} & \mathrm{\{1,1\}} & \mathrm{\{2,1\}} & \mathrm{\{3,1\}} & \mathrm{\{4,1\}} & \mathrm{\{1,2\}} & \mathrm{\{2,2\}} & \mathrm{\{3,2\}} & \mathrm{\{4,2\}} & \mathrm{\{1,3\}} & \mathrm{\{2,3\}} & \mathrm{\{3,3\}} & \mathrm{\{4,3\}} & \mathrm{\{1,4\}} & \mathrm{\{2,4\}} & \mathrm{\{3,4\}} & \mathrm{\{4,4\}}\\
 {\{j,k\}} &  &  &  &  &  &  &  &  &  &  &  &  &  &  &  &\\
\hline
 \mathrm{\{1,1\}}&   1   &   0    &   0   &   1  &   0  &   0  &   0  &   0  &   0  &   0    &   0  & 0   &   1   &   0    &   0   &   1  \\

\mathrm{\{2,1\}} &   0   &   0    &   0   &   0  &   1  &   0  &   0  &   1  &  -1  &   0    &   0  & -1  &   0   &   0    &   0   &   0  \\ 

\mathrm{\{3,1\}}   &   0    &   1    &   -1  &   0  &   0  &   0  &   0  &   0  &   0  &   0   &   0  &  0  &  0    &   1    &   -1  &   0  \\

\mathrm{\{4,1\}}   &  0    &   0    &    0  &   0  &   0  &   1  &  -1  &   0  &   0  &  -1    &   1  &   0 &  0    &   0    &   0  &   0  \\
\hline
\mathrm{\{1,2\}}   &  0    &   0    &    0  &   0  &   1  &   0  &   0  &   1  &   1  &   0    &   0  &   1 &  0    &   0    &    0  &   0  \\

\mathrm{\{2,2\}}   &  1    &   0    &    0  &   1  &   0  &   0  &   0  &   0  &   0  &   0    &   0  &   0 &  0    &   0    &    0  &   0  \\

\mathrm{\{3,2\}}   &  0    &   0    &   0  &   0  &   0  &   1  &   -1  &  0  &   0  &   1    &   -1  &  0 &  0    &   0    &    0  &   0  \\

\mathrm{\{4,2\}}   &  0    &   1    &   -1  &   0  &   0  &   0  &   0  &   0  &   0  &   0     &   0  &   0 &  0    &   -1   &    1  &   0  \\
 \hline

\mathrm{\{1,3\}}   &  0    &   1    &    1  &   0  &   0  &   0  &   0  &   0  &   0   &  0     &   0  &   0 &  0    &  -1    &    1  &   0 \\

\mathrm{\{2,3\}}   &  0    &   0    &    0  &   0  &   0  &   1  &   1  &   0  &   0   & -1    &  -1  &   0 &  0    &   0    &    0  &   0 \\

\mathrm{\{3,3\}}   &  1    &   0    &    0  &  -1  &   0  &   0  &   0  &   0  &   0   &  0     &   0  &   0 &  1    &   0    &    0  &   -1 \\

\mathrm{\{4,3\}}   &  0    &   0    &    0  &   0  &   1  &   0  &   0  &  -1  &  -1   &  0     &  0   &   1 &  0    &   0    &    0  &   0  \\
\hline
\mathrm{\{1,4\}}   &  0    &   0    &    0  &   0  &   0  &   1  &   1  &   0  &   0   &  1     &  1   &  0  &  0    &   0    &    0  &   0  \\

\mathrm{\{2,4\}}   &  0    &   1    &    1  &   0  &   0  &   0  &   0  &   0  &   0   &  0     &  0   &  0  &  0    &  -1    &   -1  &   0  \\
 
\mathrm{\{3,4\}}   &  0    &   0    &    0  &   0  &   1  &   0  &   0  &  -1  &   1   &  0     &  0   &  -1 &  0    &   0    &    0  &   0 \\
 
\mathrm{\{4,4\}}   &  1    &   0    &    0  &  -1  &   0  &   0  &   0  &   0  &   0   &  0     &  0   &  0  &  -1   &   0    &    0  &   1 \\
 \hline
 
\end{array}
    $
\end{table}